%%%
%     First, a big macro file (apple.tex, a modified version of phyzzx)
%%%

%     This is setup.tex and phyzzx.tex ,merged , and with
%     hoffset and voffset changed to suit the Apple laserwriter
%     November 88.
%%%%%%%%%%%%%%%%%%%%%%%%%%%%%%%%%%%%%%%%%%%%%%%%%%%%%%%%%%%%%%%%%%%%%%%%%
% % % % % % % % % % % % % % % % % % % % % % % % % % % % % % % % % % % %
%%%   This is PHYZZX macro package.   % % % % % % % % % % % % % % % % %
%% % % % % % % % % % % % % % % % % % % % % % % % % % % % % % % % % % % %
%%%  This version of PHYZZX should be used with Version 1.0 of TEX  % %
%% % % % % % % % % % % % % % % % % % % % % % % % % % % % % % % % % % % %
%%%   Do not "\input phyzzx" unless you preload or "\input" PLAIN.  % %
%% % % % % % % % % % % % % % % % % % % % % % % % % % % % % % % % % % % %
%%%   To preload both PLAIN and PHYZZX, begin your file with    % % % %
%%%  a line "%macropackage=phyzzx" instead of "\input phyzzx".  % % % %
%% % % % % % % % % % % % % % % % % % % % % % % % % % % % % % % % % % % %
%%%%%%%%%%%%%%%%%%%%%%%%%%%%%%%%%%%%%%%%%%%%%%%%%%%%%%%%%%%%%%%%%%%%%%%%
%%%%%%%  Created by Vadim Kaplunovsky in June 1984.   %%%%%%%%%%%%%%%%%%
% % % % % % % % % % % % % % % % % % % % % % % % % % % % % % % % % % % %
%%%%%%%%%%%%  Latest update/debug: May 29, 1985   %%%%%%%%%%%%%%%%%%%%%%
%%%%%%%%%%%%%%%%%%%%%%%%%%%%%%%%%%%%%%%%%%%%%%%%%%%%%%%%%%%%%%%%%%%%%%%%
%
\expandafter\ifx\csname phyzzx\endcsname\relax\else
 \errhelp{Hit <CR> and go ahead.}
 \errmessage{PHYZZX macros are already loaded or input. }
 \endinput \fi
\catcode`\@=11 % This allows us to modify PLAIN macros.
%
%%%%%%%%%%%%%%%%%%%%%%%%%%%%%%%%%%%%%%%%%%%%%%%%%%%%%%%%%%%%%%%%%%%%%%%%
%
%   I begin with fonts.
%
\font\seventeenrm=cmr17
\font\fourteenrm=cmr12 scaled\magstep1
\font\twelverm=cmr12
\font\ninerm=cmr9            \font\sixrm=cmr6
%
%\font\seventeenbf=cmbx10 scaled\magstep3
\font\fourteenbf=cmbx10 scaled\magstep2
\font\twelvebf=cmbx12
\font\ninebf=cmbx9            \font\sixbf=cmbx6
%
%\font\seventeeni=cmmi10 scaled\magstep3     \skewchar\seventeeni='177
\font\fourteeni=cmmi10 scaled\magstep2      \skewchar\fourteeni='177
\font\twelvei=cmmi12			        \skewchar\twelvei='177
\font\ninei=cmmi9                           \skewchar\ninei='177
\font\sixi=cmmi6                            \skewchar\sixi='177
%
%\font\seventeensy=cmsy10 scaled\magstep3    \skewchar\seventeensy='60
\font\fourteensy=cmsy10 scaled\magstep2     \skewchar\fourteensy='60
\font\twelvesy=cmsy10 scaled\magstep1	    \skewchar\twelvesy='60
\font\ninesy=cmsy9                          \skewchar\ninesy='60
\font\sixsy=cmsy6                           \skewchar\sixsy='60
%
%\font\seventeenex=cmex10 scaled\magstep3
\font\fourteenex=cmex10 scaled\magstep2
\font\twelveex=cmex10 scaled\magstep1
%\font\elevenex=cmex10 scaled\magstephalf
%
%\font\seventeensl=cmsl10 scaled\magstep3
\font\fourteensl=cmsl12 scaled\magstep1
\font\twelvesl=cmsl12
\font\ninesl=cmsl9
%
%\font\seventeenit=cmti10 scaled\magstep3
\font\fourteenit=cmti12 scaled\magstep1
\font\twelveit=cmti12
\font\nineit=cmti9
\font\fourteentt=cmtt10 scaled\magstep2
\font\twelvett=cmtt12
\font\fourteencp=cmcsc10 scaled\magstep2
\font\twelvecp=cmcsc10 scaled\magstep1
\font\tencp=cmcsc10
\newfam\cpfam
\newdimen\b@gheight		\b@gheight=12pt
\newcount\f@ntkey		\f@ntkey=0
\def\f@m{\afterassignment\samef@nt\f@ntkey=}
\def\samef@nt{\fam=\f@ntkey \the\textfont\f@ntkey\relax}
\def\rm{\f@m0 }
\def\mit{\f@m1 }         
\def\cal{\f@m2 }
\def\it{\f@m\itfam}
\def\sl{\f@m\slfam}
\def\bf{\f@m\bffam}
\def\tt{\f@m\ttfam}
\def\caps{\f@m\cpfam}
\def\fourteenpoint{\relax
    \textfont0=\fourteenrm          \scriptfont0=\tenrm
      \scriptscriptfont0=\sevenrm
    \textfont1=\fourteeni           \scriptfont1=\teni
      \scriptscriptfont1=\seveni
    \textfont2=\fourteensy          \scriptfont2=\tensy
      \scriptscriptfont2=\sevensy
    \textfont3=\fourteenex          \scriptfont3=\twelveex
      \scriptscriptfont3=\tenex
    \textfont\itfam=\fourteenit     \scriptfont\itfam=\tenit
    \textfont\slfam=\fourteensl     \scriptfont\slfam=\tensl
    \textfont\bffam=\fourteenbf     \scriptfont\bffam=\tenbf
      \scriptscriptfont\bffam=\sevenbf
    \textfont\ttfam=\fourteentt
    \textfont\cpfam=\fourteencp
    \samef@nt
    \b@gheight=14pt
    \setbox\strutbox=\hbox{\vrule height 0.85\b@gheight
				depth 0.35\b@gheight width\z@ }}
\def\twelvepoint{\relax
    \textfont0=\twelverm          \scriptfont0=\ninerm
      \scriptscriptfont0=\sixrm
    \textfont1=\twelvei           \scriptfont1=\ninei
      \scriptscriptfont1=\sixi
    \textfont2=\twelvesy           \scriptfont2=\ninesy
      \scriptscriptfont2=\sixsy
    \textfont3=\twelveex          \scriptfont3=\tenex
      \scriptscriptfont3=\tenex
    \textfont\itfam=\twelveit     \scriptfont\itfam=\nineit
    \textfont\slfam=\twelvesl     \scriptfont\slfam=\ninesl
    \textfont\bffam=\twelvebf     \scriptfont\bffam=\ninebf
      \scriptscriptfont\bffam=\sixbf
    \textfont\ttfam=\twelvett
    \textfont\cpfam=\twelvecp
    \samef@nt
    \b@gheight=12pt
    \setbox\strutbox=\hbox{\vrule height 0.85\b@gheight
				depth 0.35\b@gheight width\z@ }}
\def\tenpoint{\relax
    \textfont0=\tenrm          \scriptfont0=\sevenrm
      \scriptscriptfont0=\fiverm
    \textfont1=\teni           \scriptfont1=\seveni
      \scriptscriptfont1=\fivei
    \textfont2=\tensy          \scriptfont2=\sevensy
      \scriptscriptfont2=\fivesy
    \textfont3=\tenex          \scriptfont3=\tenex
      \scriptscriptfont3=\tenex
    \textfont\itfam=\tenit     \scriptfont\itfam=\seveni
    \textfont\slfam=\tensl     \scriptfont\slfam=\sevenrm
    \textfont\bffam=\tenbf     \scriptfont\bffam=\sevenbf
      \scriptscriptfont\bffam=\fivebf
    \textfont\ttfam=\tentt
    \textfont\cpfam=\tencp
    \samef@nt
    \b@gheight=10pt
    \setbox\strutbox=\hbox{\vrule height 0.85\b@gheight
				depth 0.35\b@gheight width\z@ }}
%
%%%%%%%%%%%%%%%%%%%%%%%%%%%%%%%%%%%%%%%%%%%%%%%%%%%%%%%%%%%%%%%%%%%%%%%%
%
%   Next, I define basic spacing parameters.
%
\normalbaselineskip = 20pt plus 0.2pt minus 0.1pt
\normallineskip = 1.5pt plus 0.1pt minus 0.1pt
\normallineskiplimit = 1.5pt
\newskip\normaldisplayskip
\normaldisplayskip = 20pt plus 5pt minus 10pt
\newskip\normaldispshortskip
\normaldispshortskip = 6pt plus 5pt
\newskip\normalparskip
\normalparskip = 6pt plus 2pt minus 1pt
\newskip\skipregister
\skipregister = 5pt plus 2pt minus 1.5pt
\newif\ifsingl@    \newif\ifdoubl@
\newif\iftwelv@    \twelv@true
\def\singlespace{\singl@true\doubl@false\spaces@t}
\def\doublespace{\singl@false\doubl@true\spaces@t}
\def\normalspace{\singl@false\doubl@false\spaces@t}
\def\Tenpoint{\tenpoint\twelv@false\spaces@t}
\def\Twelvepoint{\twelvepoint\twelv@true\spaces@t}
\def\spaces@t{\relax
      \iftwelv@ \ifsingl@\subspaces@t3:4;\else\subspaces@t1:1;\fi
       \else \ifsingl@\subspaces@t3:5;\else\subspaces@t4:5;\fi \fi
      \ifdoubl@ \multiply\baselineskip by 5
         \divide\baselineskip by 4 \fi }
\def\subspaces@t#1:#2;{
      \baselineskip = \normalbaselineskip
      \multiply\baselineskip by #1 \divide\baselineskip by #2
      \lineskip = \normallineskip
      \multiply\lineskip by #1 \divide\lineskip by #2
      \lineskiplimit = \normallineskiplimit
      \multiply\lineskiplimit by #1 \divide\lineskiplimit by #2
      \parskip = \normalparskip
      \multiply\parskip by #1 \divide\parskip by #2
      \abovedisplayskip = \normaldisplayskip
      \multiply\abovedisplayskip by #1 \divide\abovedisplayskip by #2
      \belowdisplayskip = \abovedisplayskip
      \abovedisplayshortskip = \normaldispshortskip
      \multiply\abovedisplayshortskip by #1
        \divide\abovedisplayshortskip by #2
      \belowdisplayshortskip = \abovedisplayshortskip
      \advance\belowdisplayshortskip by \belowdisplayskip
      \divide\belowdisplayshortskip by 2
      \smallskipamount = \skipregister
      \multiply\smallskipamount by #1 \divide\smallskipamount by #2
      \medskipamount = \smallskipamount \multiply\medskipamount by 2
      \bigskipamount = \smallskipamount \multiply\bigskipamount by 4 }
\def\normalbaselines{ \baselineskip=\normalbaselineskip
   \lineskip=\normallineskip \lineskiplimit=\normallineskip
   \iftwelv@\else \multiply\baselineskip by 4 \divide\baselineskip by 5
     \multiply\lineskiplimit by 4 \divide\lineskiplimit by 5
     \multiply\lineskip by 4 \divide\lineskip by 5 \fi }
\Twelvepoint  % That's the default
\interlinepenalty=50
\interfootnotelinepenalty=5000
\predisplaypenalty=9000
\postdisplaypenalty=500
\hfuzz=1pt
\vfuzz=0.2pt
\voffset=0pt
\dimen\footins=8 truein
%
%%%%%%%%%%%%%%%%%%%%%%%%%%%%%%%%%%%%%%%%%%%%%%%%%%%%%%%%%%%%%%%%%%%%%%%%
%
%   Next, I define output routines, footnotes & related stuff.
%
\def\pagecontents{
   \ifvoid\topins\else\unvbox\topins\vskip\skip\topins\fi
   \dimen@ = \dp255 \unvbox255
   \ifvoid\footins\else\vskip\skip\footins\footrule\unvbox\footins\fi
   \ifr@ggedbottom \kern-\dimen@ \vfil \fi }
\def\makeheadline{\vbox to 0pt{ \skip@=\topskip
      \advance\skip@ by -12pt \advance\skip@ by -2\normalbaselineskip
      \vskip\skip@ \line{\vbox to 12pt{}\the\headline} \vss
      }\nointerlineskip}
\def\makefootline{\baselineskip = 1.5\normalbaselineskip
                 \line{\the\footline}}
\newif\iffrontpage
\newif\ifletterstyle
\newif\ifp@genum
\def\nopagenumbers{\p@genumfalse}
\def\pagenumbers{\p@genumtrue}
\pagenumbers
\newtoks\paperheadline
\newtoks\letterheadline
\newtoks\paperfootline
\newtoks\letterfootline
\newtoks\letterinfo
\newtoks\Letterinfo
\newtoks\date
\footline={\ifletterstyle\the\letterfootline\else\the\paperfootline\fi}
\paperfootline={\hss\iffrontpage\else\ifp@genum\tenrm\folio\hss\fi\fi}
\letterfootline={\iffrontpage\LETTERFOOT\else\hfil\fi}
\Letterinfo={\hfil}
\letterinfo={\hfil}
\def\LETTERFOOT{\hfil} %\ninerm PASADENA, CALIFORNIA 91125 TELEPHONE:
%(818)356-6686\hfil}
%
\def\LETTERHEAD{\vtop{\baselineskip=20pt\hbox to
\hsize{\hfil\seventeenrm\strut 
CALIFORNIA INSTITUTE OF TECHNOLOGY \hfil}
\hbox to \hsize{\hfil\ninerm\strut
CHARLES C. LAURITSEN LABORATORY OF HIGH ENERGY PHYSICS \hfil}
\hbox to \hsize{\hfil\ninerm\strut
PASADENA, CALIFORNIA 91125 \hfil}}}
\headline={\ifletterstyle\the\letterheadline\else\the\paperheadline\fi}
\paperheadline={\hfil}
\letterheadline{\iffrontpage \LETTERHEAD\else
    \rm \ifp@genum \hfil \folio\hfil\fi\fi}
\def\monthname{\relax\ifcase\month 0/\or January\or February\or
   March\or April\or May\or June\or July\or August\or September\or
   October\or November\or December\else\number\month/\fi}
\def\today{\monthname\ \number\day, \number\year}
\date={\today}
\countdef\pageno=1      \countdef\pagen@=0
\countdef\pagenumber=1  \pagenumber=1
\def\advancepageno{\global\advance\pagen@ by 1
   \ifnum\pagenumber<0 \global\advance\pagenumber by -1
    \else\global\advance\pagenumber by 1 \fi \global\frontpagefalse }
\def\folio{\ifnum\pagenumber<0 \romannumeral-\pagenumber
           \else \number\pagenumber \fi }
\def\footrule{\dimen@=\prevdepth\nointerlineskip
   \vbox to 0pt{\vskip -0.25\baselineskip \hrule width 0.35\hsize \vss}
   \prevdepth=\dimen@ }
\newtoks\foottokens
\foottokens={}
\newdimen\footindent
\footindent=24pt
\def\vfootnote#1{\insert\footins\bgroup  
   \interlinepenalty=\interfootnotelinepenalty \floatingpenalty=20000
   \singl@true\doubl@false\Tenpoint
   \splittopskip=\ht\strutbox \boxmaxdepth=\dp\strutbox
   \leftskip=\footindent \rightskip=\z@skip
   \parindent=0.5\footindent \parfillskip=0pt plus 1fil
   \spaceskip=\z@skip \xspaceskip=\z@skip
   \the\foottokens
   \Textindent{$ #1 $}\footstrut\futurelet\next\fo@t}
\def\Textindent#1{\noindent\llap{#1\enspace}\ignorespaces}
\def\footnote#1{\attach{#1}\vfootnote{#1}}

\let\footsymbol=\star
\newcount\lastf@@t           \lastf@@t=-1
\newcount\footsymbolcount    \footsymbolcount=0
\newif\ifPhysRev
\def\bumpfootsymbolcount{\relax
   \iffrontpage \bumpfootsymbolNP \else \advance\lastf@@t by 1
     \ifPhysRev \bumpfootsymbolPR \else \bumpfootsymbolNP \fi \fi
   \global\lastf@@t=\pagen@ }
\def\bumpfootsymbolNP{\ifnum\footsymbolcount <0 \global\footsymbolcount =0 \fi
    \ifnum\lastf@@t<\pagen@ \global\footsymbolcount=0
     \else \global\advance\footsymbolcount by 1 \fi }
\def\bumpfootsymbolPR{\ifnum\footsymbolcount >0 \global\footsymbolcount =0 \fi
      \global\advance\footsymbolcount by -1 }
\def\fd@f#1 {\xdef\footsymbol{\mathchar"#1 }}
\def\generatefootsymbol{\ifcase\footsymbolcount \fd@f 13F \or \fd@f 279
	\or \fd@f 27A \or \fd@f 278 \or \fd@f 27B \else
	\ifnum\footsymbolcount <0 \fd@f{023 \number-\footsymbolcount }
	 \else \fd@f 203 {\loop \ifnum\footsymbolcount >5
		\fd@f{203 \footsymbol } \advance\footsymbolcount by -1
		\repeat }\fi \fi }

\def\nonfrenchspacing{\sfcode`\.=3001 \sfcode`\!=3000 \sfcode`\?=3000
	\sfcode`\:=2000 \sfcode`\;=1500 \sfcode`\,=1251 }
\nonfrenchspacing
\newdimen\d@twidth
{\setbox0=\hbox{s.} \global\d@twidth=\wd0 \setbox0=\hbox{s}
	\global\advance\d@twidth by -\wd0 }
\def\removehglue{\loop \unskip \ifdim\lastskip >\z@ \repeat }
\def\roll@ver#1{\removehglue \nobreak \count255 =\spacefactor \dimen@=\z@
	\ifnum\count255 =3001 \dimen@=\d@twidth \fi
	\ifnum\count255 =1251 \dimen@=\d@twidth \fi
    \iftwelv@ \kern-\dimen@ \else \kern-0.83\dimen@ \fi
   #1\spacefactor=\count255 }
\def\step@ver#1{\relax \ifmmode #1\else \ifhmode
	\roll@ver{${}#1$}\else {\setbox0=\hbox{${}#1$}}\fi\fi }
\def\attach#1{\step@ver{\strut^{\mkern 2mu #1} }}
%
%%%%%%%%%%%%%%%%%%%%%%%%%%%%%%%%%%%%%%%%%%%%%%%%%%%%%%%%%%%%%%%%%%%%%%%%
%
%   Here come chapter, section, subsection & appendix macros.
%
\newcount\chapternumber      \chapternumber=0
\newcount\sectionnumber      \sectionnumber=0
\newcount\equanumber         \equanumber=0
\let\chapterlabel=\relax
\let\sectionlabel=\relax
\newtoks\chapterstyle        \chapterstyle={\Number}
\newtoks\sectionstyle        \sectionstyle={\chapterlabel\Number}
\newskip\chapterskip         \chapterskip=\bigskipamount
\newskip\sectionskip         \sectionskip=\medskipamount
\newskip\headskip            \headskip=8pt plus 3pt minus 3pt
\newdimen\chapterminspace    \chapterminspace=15pc
\newdimen\sectionminspace    \sectionminspace=10pc
\newdimen\referenceminspace  \referenceminspace=25pc
\def\chapterreset{\global\advance\chapternumber by 1
   \ifnum\equanumber<0 \else\global\equanumber=0\fi
   \sectionnumber=0 \makechapterlabel}
\def\makechapterlabel{\let\sectionlabel=\relax
   \xdef\chapterlabel{\the\chapterstyle{\the\chapternumber}.}}
\def\alphabetic#1{\count255='140 \advance\count255 by #1\char\count255}
\def\Alphabetic#1{\count255='100 \advance\count255 by #1\char\count255}
\def\Roman#1{\uppercase\expandafter{\romannumeral #1}}
\def\roman#1{\romannumeral #1}
\def\Number#1{\number #1}
\def\BLANC#1{}
\def\titlestyle#1{\par\begingroup \interlinepenalty=9999
     \leftskip=0.02\hsize plus 0.23\hsize minus 0.02\hsize
     \rightskip=\leftskip \parfillskip=0pt
     \hyphenpenalty=9000 \exhyphenpenalty=9000
     \tolerance=9999 \pretolerance=9000
     \spaceskip=0.333em \xspaceskip=0.5em
     \iftwelv@\fourteenpoint\else\twelvepoint\fi
   \noindent #1\par\endgroup }
\def\spacecheck#1{\dimen@=\pagegoal\advance\dimen@ by -\pagetotal
   \ifdim\dimen@<#1 \ifdim\dimen@>0pt \vfil\break \fi\fi}
\def\TableOfContentEntry#1#2#3{\relax}
\def\chapter#1{\par \penalty-300 \vskip\chapterskip
   \spacecheck\chapterminspace
   \chapterreset \titlestyle{\chapterlabel\ #1}
   \TableOfContentEntry c\chapterlabel{#1}
   \nobreak\vskip\headskip \penalty 30000
   \wlog{\string\chapter\space \chapterlabel} }

\def\section#1{\par \ifnum\the\lastpenalty=30000\else
   \penalty-200\vskip\sectionskip \spacecheck\sectionminspace\fi
   \global\advance\sectionnumber by 1
   \xdef\sectionlabel{\the\sectionstyle\the\sectionnumber}
   \wlog{\string\section\space \sectionlabel}
   \TableOfContentEntry s\sectionlabel{#1}
   \noindent {\caps\enspace\sectionlabel\quad #1}\par
   \nobreak\vskip\headskip \penalty 30000 }
\def\subsection#1{\par
   \ifnum\the\lastpenalty=30000\else \penalty-100\smallskip \fi
   \noindent\undertext{#1}\enspace \vadjust{\penalty5000}}

\def\undertext#1{\vtop{\hbox{#1}\kern 1pt \hrule}}
\def\ack{\par\penalty-100\medskip \spacecheck\sectionminspace
   \line{\fourteenrm\hfil ACKNOWLEDGEMENTS\hfil}\nobreak\vskip\headskip }
\def\APPENDIX#1#2{\par\penalty-300\vskip\chapterskip
   \spacecheck\chapterminspace \chapterreset \xdef\chapterlabel{#1}
   \titlestyle{APPENDIX #2} \nobreak\vskip\headskip \penalty 30000
   \TableOfContentEntry a{#1}{#2}
   \wlog{\string\Appendix\ \chapterlabel} }
\def\Appendix#1{\APPENDIX{#1}{#1}}
\def\appendix{\APPENDIX{A}{}}
\def\unnumberedchapters{\let\makechapterlabel=\relax \let\chapterlabel=\relax
   \sectionstyle={\BLANC}\let\sectionlabel=\relax \sequentialequations }
%
%%%%%%%%%%%%%%%%%%%%%%%%%%%%%%%%%%%%%%%%%%%%%%%%%%%%%%%%%%%%%%%%%%%%%%%%
%
%   Here come macros for equation numbering.
%
\def\eqname#1{\relax \ifnum\equanumber<0
     \xdef#1{{\noexpand\rm(\number-\equanumber)}}%
       \global\advance\equanumber by -1
    \else \global\advance\equanumber by 1
      \xdef#1{{\noexpand\rm(\chapterlabel\number\equanumber)}} \fi #1}

\def\eqn{\eqno\eqname}

\def\eqinsert#1{\noalign{\dimen@=\prevdepth \nointerlineskip
   \setbox0=\hbox to\displaywidth{\hfil #1}
   \vbox to 0pt{\kern 0.5\baselineskip\hbox{$\!\box0\!$}\vss}
   \prevdepth=\dimen@}}
%

%
%%%%%%%%%%%%%%%%%%%%%%%%%%%%%%%%%%%%%%%%%%%%%%%%%%%%%%%%%%%%%%%%%%%%%%%%
%   Here come items and lists
%
\def\GENITEM#1;#2{\par \hangafter=0 \hangindent=#1
    \Textindent{$ #2 $}\ignorespaces}
\outer\def\newitem#1=#2;{\gdef#1{\GENITEM #2;}}
\newdimen\itemsize                \itemsize=30pt
\newitem\item=1\itemsize;
\newitem\sitem=1.75\itemsize;     
\newitem\ssitem=2.5\itemsize;     
\outer\def\newlist#1=#2&#3&#4;{\toks0={#2}\toks1={#3}%
   \count255=\escapechar \escapechar=-1
   \alloc@0\list\countdef\insc@unt\listcount     \listcount=0
   \edef#1{\par
      \countdef\listcount=\the\allocationnumber
      \advance\listcount by 1
      \hangafter=0 \hangindent=#4
      \Textindent{\the\toks0{\listcount}\the\toks1}}
   \expandafter\expandafter\expandafter
    \edef\c@t#1{begin}{\par
      \countdef\listcount=\the\allocationnumber \listcount=1
      \hangafter=0 \hangindent=#4
      \Textindent{\the\toks0{\listcount}\the\toks1}}
   \expandafter\expandafter\expandafter
    \edef\c@t#1{con}{\par \hangafter=0 \hangindent=#4 \noindent}
   \escapechar=\count255}
\def\c@t#1#2{\csname\string#1#2\endcsname}
\newlist\point=\Number&.&1.0\itemsize;
\newlist\subpoint=(\alphabetic&)&1.75\itemsize;
\newlist\subsubpoint=(\roman&)&2.5\itemsize;
%

%
%%%%%%%%%%%%%%%%%%%%%%%%%%%%%%%%%%%%%%%%%%%%%%%%%%%%%%%%%%%%%%%%%%%%%%%%
%
%   Here come macros for references, figures & tables.
%
% % % % % % % % % % % % % % % % % % % % % % % % % % % % % % % % % % % %
%%  First, references.
%
\newcount\referencecount     \referencecount=0
\newcount\lastrefsbegincount \lastrefsbegincount=0
\newif\ifreferenceopen       \newwrite\referencewrite
\newif\ifrw@trailer
\newdimen\refindent     \refindent=30pt
\def\NPrefmark#1{\attach{\scriptscriptstyle [ #1 ] }}
\let\PRrefmark=\attach
\def\refmark#1{\relax\ifPhysRev\PRrefmark{#1}\else\NPrefmark{#1}\fi}
\def\refend@{\refmark{\number\referencecount}}
\def\refend{\refend@{}\space }
\def\refsend{\refmark{\count255=\referencecount
   \advance\count255 by-\lastrefsbegincount
   \ifcase\count255 \number\referencecount
   \or \number\lastrefsbegincount,\number\referencecount
   \else \number\lastrefsbegincount-\number\referencecount \fi}\space }
\def\refitem#1{\par \hangafter=0 \hangindent=\refindent \Textindent{#1}}
\def\Ref{\rw@trailertrue\REF}
\def\ref{\Ref\?}

\def\REF#1{\r@fstart{#1}%
   \rw@begin{\the\referencecount.}\rw@end}
\def\REFS#1{\r@fstart{#1}%
   \lastrefsbegincount=\referencecount
   \rw@begin{\the\referencecount.}\rw@end}
\def\r@fstart#1{\chardef\rw@write=\referencewrite \let\rw@ending=\refend@
   \ifreferenceopen \else \global\referenceopentrue
   \immediate\openout\referencewrite=referenc.txa
   \toks0={\catcode`\^^M=10}\immediate\write\rw@write{\the\toks0} \fi
   \global\advance\referencecount by 1 \xdef#1{\the\referencecount}}
{\catcode`\^^M=\active %
 \gdef\rw@begin#1{\immediate\write\rw@write{\noexpand\refitem{#1}}%
   \begingroup \catcode`\^^M=\active \let^^M=\relax}%
 \gdef\rw@end#1{\rw@@end #1^^M\rw@terminate \endgroup%
   \ifrw@trailer\rw@ending\global\rw@trailerfalse\fi }%
 \gdef\rw@@end#1^^M{\toks0={#1}\immediate\write\rw@write{\the\toks0}%
   \futurelet\n@xt\rw@test}%
 \gdef\rw@test{\ifx\n@xt\rw@terminate \let\n@xt=\relax%
       \else \let\n@xt=\rw@@end \fi \n@xt}%
}
\let\rw@ending=\relax
\let\rw@terminate=\relax
\let\splitout=\relax
\def\Closeout#1{\toks0={\catcode`\^^M=5}\immediate\write#1{\the\toks0}%
   \immediate\closeout#1}
%
% % % % % % % % % % % % % % % % % % % % % % % % % % % % % % % % % % % %
%%  Next, figure captions and table captions.
%
\newcount\figurecount     \figurecount=0
\newcount\tablecount      \tablecount=0
\newif\iffigureopen       \newwrite\figurewrite
\newif\iftableopen        \newwrite\tablewrite
\def\FIG#1{\f@gstart{#1}%
   \rw@begin{\the\figurecount)}\rw@end}

\def\Fig{\rw@trailertrue\def\rw@ending{Fig.~\?}\FIG\?}
\def\fig{\rw@trailertrue\def\rw@ending{fig.~\?}\FIG\?}
\def\TABLE#1{\T@Bstart{#1}%
   \rw@begin{\the\tableecount:}\rw@end}
\def\Table{\rw@trailertrue\def\rw@ending{Table~\?}\TABLE\?}
\def\f@gstart#1{\chardef\rw@write=\figurewrite
   \iffigureopen \else \global\figureopentrue
   \immediate\openout\figurewrite=figures.txa
   \toks0={\catcode`\^^M=10}\immediate\write\rw@write{\the\toks0} \fi
   \global\advance\figurecount by 1 \xdef#1{\the\figurecount}}
\def\T@Bstart#1{\chardef\rw@write=\tablewrite
   \iftableopen \else \global\tableopentrue
   \immediate\openout\tablewrite=tables.txa
   \toks0={\catcode`\^^M=10}\immediate\write\rw@write{\the\toks0} \fi
   \global\advance\tablecount by 1 \xdef#1{\the\tablecount}}
%

%
% % % % % % % % % % % % % % % % % % % % % % % % % % % % % % % % % % % %
%%  Finally, inserted figures.
%
%\newread\figureread                                     %% That's
%\def\g@tfigure#1#2 {\openin\figureread #2.fig           %% an example
%   \ifeof\figureread \errhelp=\disabledfigures          %% of \g@tfigure
%     \errmessage{No such file: #2.fig}\let#1=\relax \else
%    \read\figureread to\y@p \read\figureread to\y@p     %%
%    \read\figureread to\x@p \read\figureread to\y@m     %% See LOCPHYX.TEX
%    \read\figureread to\x@m \closein\figureread         %% file for the
%    \xdef#1{\hbox{\kern-\x@m truein \vbox{\kern-\y@m truein
%      \hbox to \x@p truein{\vbox to \y@p truein{        %% actual definition.
%        \special{pos,inc=#2.fig}\vss }\hss }}}}\fi }    %%
%
\def\getfigure#1{\global\advance\figurecount by 1
   \xdef#1{\the\figurecount}\count255=\escapechar \escapechar=-1
   \edef\n@xt{\noexpand\g@tfigure\csname\string#1Body\endcsname}%
   \escapechar=\count255 \n@xt }
\def\g@tfigure#1#2 {\errhelp=\disabledfigures \let#1=\relax
   \errmessage{\string\getfigure\space disabled}}
\newhelp\disabledfigures{ Empty figure of zero size assumed.}
\def\figinsert#1{\midinsert\Tenpoint\medskip
   \count255=\escapechar \escapechar=-1
   \edef\n@xt{\csname\string#1Body\endcsname}
   \escapechar=\count255 \centerline{\n@xt}
   \bigskip\narrower\narrower
   \noindent{\it Figure}~#1.\quad }
%
%%%%%%%%%%%%%%%%%%%%%%%%%%%%%%%%%%%%%%%%%%%%%%%%%%%%%%%%%%%%%%%%%%%%%%%%
%
%   Here come macros for memos & letters.
%
\def\masterreset{\global\pagenumber=1 \global\chapternumber=0
   \global\equanumber=0 \global\sectionnumber=0
   \global\referencecount=0 \global\figurecount=0 \global\tablecount=0 }
\def\FRONTPAGE{\ifvoid255\else\vfill\penalty-20000\fi
      \masterreset\global\frontpagetrue
      \global\lastf@@t=0 \global\footsymbolcount=0}

\def\paperstyle{\letterstylefalse\normalspace\papersize}
\def\letterstyle{\letterstyletrue\singlespace\lettersize}
%  old hoffset = 2 truepc         November 1988
\def\papersize{\hsize=35 truepc\vsize=50 truepc\hoffset=-2.51688 truepc
               \skip\footins=\bigskipamount}
%  old hoffset = 1.3 truein  old voffset = 1.25 truein  November 1988
\def\lettersize{\hsize=5.5 truein\vsize=8.25 truein\hoffset=.4875 truein
	\voffset=.3125 truein
   \skip\footins=\smallskipamount \multiply\skip\footins by 3 }
\paperstyle   %  This is the default
%
% % % % % % % % % % % % % % % % % % % % % % % % % % % % % % % % % % % %
%
\def\MEMO{\letterstyle \letterinfo={\hfil } \let\rule=\memorule
	\FRONTPAGE \memohead }
\let\memohead=\relax

\def\memit@m#1{\smallskip \hangafter=0 \hangindent=1in
      \Textindent{\caps #1}}
\def\subject{\memit@m{Subject:}}
\def\topic{\memit@m{Topic:}}
\def\from{\memit@m{From:}}
\def\to{\relax \ifmmode \rightarrow \else \memit@m{To:}\fi }
\def\memorule{\medskip\hrule height 1pt\bigskip}
\newwrite\labelswrite
\newtoks\rw@toks

\def\addressee#1{\null\vskip .5truein\line{
\hskip 0.5\hsize minus 0.5\hsize\the\date\hfil}\bigskip
   \ialign to\hsize{\strut ##\hfil\tabskip 0pt plus \hsize \cr #1\crcr}
   \writelabel{#1}\medskip\par\noindent}
\def\rwl@begin#1\cr{\rw@toks={#1\crcr}\relax
   \immediate\write\labelswrite{\the\rw@toks}\futurelet\n@xt\rwl@next}
\def\rwl@next{\ifx\n@xt\rwl@end \let\n@xt=\relax
      \else \let\n@xt=\rwl@begin \fi \n@xt}
\let\rwl@end=\relax
\def\writelabel#1{\immediate\write\labelswrite{\noexpand\labelbegin}
     \rwl@begin #1\cr\rwl@end
     \immediate\write\labelswrite{\noexpand\labelend}}
\newbox\FromLabelBox

\let\labelend=\relax
\def\labelbegin#1\labelend{\setbox0=\vbox{\ialign{##\hfil\cr #1\crcr}}
     \MakeALabel }
\newtoks\FromAddress
\FromAddress={}
\def\MakeFromBox#1{\global\setbox\FromLabelBox=\vbox{\Tenpoint
     \ialign{##\hfil\cr #1\the\FromAddress\crcr}}}
\newdimen\labelwidth		\labelwidth=6in
\def\MakeALabel{\vskip 1pt \hbox{\vrule \vbox{
	\hsize=\labelwidth \hrule\bigskip
	\leftline{\hskip 1\parindent \copy\FromLabelBox}\bigskip
	\centerline{\hfil \box0 } \bigskip \hrule
	}\vrule } \vskip 1pt plus 1fil }
\newskip\signatureskip       \signatureskip=30pt
\def\signed#1{\par \penalty 9000 \medskip \dt@pfalse
  \everycr={\noalign{\ifdt@p\vskip\signatureskip\global\dt@pfalse\fi}}
  \setbox0=\vbox{\singlespace \ialign{\strut ##\hfil\crcr
   \noalign{\global\dt@ptrue}#1\crcr}}
  \line{\hskip 0.5\hsize minus 0.5\hsize \box0\hfil} \medskip }
\newbox\letterb@x
\def\lettertext{\par\unvcopy\letterb@x\par}
\def\multiletter{\setbox\letterb@x=\vbox\bgroup
      \everypar{\vrule height 1\baselineskip depth 0pt width 0pt }
      \singlespace \topskip=\baselineskip }
\def\letterend{\par\egroup}
%
%%%%%%%%%%%%%%%%%%%%%%%%%%%%%%%%%%%%%%%%%%%%%%%%%%%%%%%%%%%%%%%%%%%%%%%
%
%   Here come macros for title pages.
%
\newskip\frontpageskip
\newtoks\Pubnum
\newtoks\pubtype
\newif\ifp@bblock  \p@bblocktrue
\def\PH@SR@V{\doubl@true \baselineskip=24.1pt plus 0.2pt minus 0.1pt
             \parskip= 3pt plus 2pt minus 1pt }
\def\PHYSREV{\paperstyle\PhysRevtrue\PH@SR@V}
\def\titlepage{\FRONTPAGE\paperstyle\ifPhysRev\PH@SR@V\fi
   \ifp@bblock\p@bblock \else\hrule height\z@ \relax \fi }
\def\nopubblock{\p@bblockfalse}

\frontpageskip=12pt plus .5fil minus 2pt
\pubtype={\tensl Preliminary Version}
\Pubnum={}
\def\p@bblock{\begingroup \tabskip=\hsize minus \hsize
   \baselineskip=1.5\ht\strutbox \topspace-2\baselineskip
   \halign to\hsize{\strut ##\hfil\tabskip=0pt\crcr
       \the\Pubnum\crcr\the\date\crcr\the\pubtype\crcr}\endgroup}
\def\title#1{\vskip\frontpageskip \titlestyle{#1} \vskip\headskip }
\def\author#1{\vskip\frontpageskip\titlestyle{\twelvecp #1}\nobreak}

\def\address#1{\par\kern 5pt\titlestyle{\twelvepoint\it #1}}
\def\andaddress{\par\kern 5pt \centerline{\sl and} \address}

\def\abstract{\par\dimen@=\prevdepth \hrule height\z@ \prevdepth=\dimen@
   \vskip\frontpageskip\centerline{\fourteenrm ABSTRACT}\vskip\headskip }

%
%
%%%%%%%%%%%%%%%%%%%%%%%%%%%%%%%%%%%%%%%%%%%%%%%%%%%%%%%%%%%%%%%%%%%%%%%%
%   Miscellaneous macros
%

\def\\{\relax \ifmmode \backslash \else {\tt\char`\\}\fi }
\def\sequentialequations{\relax\if\equanumber<0\else\global\equanumber=-1\fi}

\def\journal#1&#2(#3){\unskip, \sl #1\unskip~\bf\ignorespaces #2\rm (19#3),}

\def\topspace{\hrule height 0pt depth 0pt \vskip}

\def\Buildrel#1\under#2{\mathrel{\mathop{#2}\limits_{#1}}}
\def\becomes#1{\mathchoice{\becomes@\scriptstyle{#1}}{\becomes@\scriptstyle
   {#1}}{\becomes@\scriptscriptstyle{#1}}{\becomes@\scriptscriptstyle{#1}}}
\def\becomes@#1#2{\mathrel{\setbox0=\hbox{$\m@th #1{\,#2\,}$}%
	\mathop{\hbox to \wd0 {\rightarrowfill}}\limits_{#2}}}

\let\int=\intop         \let\oint=\ointop
\def\lsim{\mathrel{\mathpalette\@versim<}}
\def\gsim{\mathrel{\mathpalette\@versim>}}
\def\@versim#1#2{\vcenter{\offinterlineskip
	\ialign{$\m@th#1\hfil##\hfil$\crcr#2\crcr\sim\crcr } }}
\def\big#1{{\hbox{$\left#1\vbox to 0.85\b@gheight{}\right.\n@space$}}}
\def\Big#1{{\hbox{$\left#1\vbox to 1.15\b@gheight{}\right.\n@space$}}}
\def\bigg#1{{\hbox{$\left#1\vbox to 1.45\b@gheight{}\right.\n@space$}}}
\def\Bigg#1{{\hbox{$\left#1\vbox to 1.75\b@gheight{}\right.\n@space$}}}
%
% % % % % % % % % % % % % % % % % % % % % % % % % % % % % % % % % % % %
%
%   Finally, some bug fixings.
%
\let\sec@nt=\sec
\def\sec{\relax\ifmmode\let\n@xt=\sec@nt\else\let\n@xt\section\fi\n@xt}
\def\obsolete#1{\message{Macro \string #1 is obsolete.}}
\def\firstsec#1{\obsolete\firstsec \section{#1}}
\def\firstsubsec#1{\obsolete\firstsubsec \subsection{#1}}
\def\thispage#1{\obsolete\thispage \global\pagenumber=#1\frontpagefalse}
\def\thischapter#1{\obsolete\thischapter \global\chapternumber=#1}
\def\REFSCON{\obsolete\REFSCON\REF}
\def\splitout{\obsolete\splitout\relax}
\def\prop{\obsolete\prop \propto }
\def\nextequation#1{\obsolete\nextequation \global\equanumber=#1
   \ifnum\the\equanumber>0 \global\advance\equanumber by 1 \fi}
\def\BOXITEM{\afterassigment\B@XITEM\setbox0=}
\def\B@XITEM{\par\hangindent\wd0 \noindent\box0 }
\def\phyzzx{PHY\setbox0=\hbox{Z}\copy0 \kern-0.5\wd0 \box0 X}
%
%%%%%%%%%%%%%%%%%%%%%%%%%%%%%%%%%%%%%%%%%%%%%%%%%%%%%%%%%%%%%%%%%%%%%%%%
%   That's about it
%
\everyjob{\xdef\today{\monthname\ \number\day, \number\year}}
        
%

%\setup

% Some macros for HEP preprints
% old hoffset=1.0 truein  November 1988
\hoffset=0.2truein
% old voffset=1.0 truein  November 1988
\voffset=0.1truein
\hsize=6truein

\def\CALT#1{\hbox to\hsize{\tenpoint \baselineskip=12pt
	\hfil\vtop{\hbox{\strut CALT-68-#1}
	\hbox{\strut DOE RESEARCH AND}
	\hbox{\strut DEVELOPMENT REPORT}}}}

\def\CALTECH{\smallskip
	\address{California Institute of Technology, Pasadena, CA 91125}}
\def\TITLE#1{\vskip 1in \centerline{\fourteenpoint #1}}
\def\AUTHOR#1{\vskip .5in \centerline{#1}}

\def\ABSTRACT#1{\vskip .5in \vfil \centerline{\twelvepoint \bf Abstract}
	#1 \vfil}

\def\sqr#1#2{{\vcenter{\hrule height.#2pt
      \hbox{\vrule width.#2pt height#1pt \kern#1pt
        \vrule width.#2pt}
      \hrule height.#2pt}}}

\def\section#1#2{
\noindent\hbox{\hbox{\bf #1}\hskip 10pt\vtop{\hsize=5in
\baselineskip=12pt \noindent \bf #2 \hfil}\hfil}
\medskip}

\def\underwig#1{	% produce a tilde below the argument
	\setbox0=\hbox{\rm \strut}
	\hbox to 0pt{$#1$\hss} \lower \ht0 \hbox{\rm \char'176}}

\def\bunderwig#1{	% produce a tilde below the argument
	\setbox0=\hbox{\rm \strut}
	\hbox to 1.5pt{$#1$\hss} \lower 12.8pt
	 \hbox{\seventeenrm \char'176}\hbox to 2pt{\hfil}}

\def\MEMO#1#2#3#4#5{
\frontpagetrue
\centerline{\tencp INTEROFFICE MEMORANDUM}
\smallskip
\centerline{\bf CALIFORNIA INSTITUTE OF TECHNOLOGY}
\centerline{\tencp Charles C. Lauritsen Laboratory of High Energy Physics} 
\bigskip
\vtop{\tenpoint \hbox to\hsize{\strut \hbox to .75in{\caps to:\hfil}
\hbox to3in{#1\hfil}
\hbox to .75in{\caps date:\hfil}\quad \the\date\hfil}
\hbox to\hsize{\strut \hbox to.75in{\caps from:\hfil}\hbox to 2in{#2\hfil}
\hbox{{\caps extension:}\quad#3\qquad{\caps mail code:\quad}#4}\hfil}
\hbox{\hbox to.75in{\caps subject:\hfil}\vtop{\parindent=0pt
\hsize=3.5in #5\hfil}}
\hbox{\strut\hfil}}}

%%%%%%%%%%%%%%%%%%%%%%%%%%%
%                         %
%   End of apple.tex      %
%                         %
%%%%%%%%%%%%%%%%%%%%%%%%%%%

%%%%%%%%%%%%%%%%%%%%%%%%%%%%%%%%%%%%%%%%%
%  begin  own macros and actual paper.  %
%%%%%%%%%%%%%%%%%%%%%%%%%%%%%%%%%%%%%%%%%

%  Use the following line only when macros are not attached.
%\input apple

\tolerance 10000

%FIGURE MACROS

\newread\figureread                                     %% That's
\def\g@tfigure#1#2 {\openin\figureread #2.fig           %% an example
   \ifeof\figureread \errhelp=\disabledfigures          %% of \g@tfigure
     \errmessage{No such file: #2.fig}\let#1=\relax \else
    \read\figureread to\y@p \read\figureread to\y@p     %%
    \read\figureread to\x@p \read\figureread to\y@m     %% See LOCPHYX.TEX
    \read\figureread to\x@m \closein\figureread         %% file for the
    \xdef#1{\hbox{\kern-\x@m truein \vbox{\kern-\y@m truein
      \hbox to \x@p truein{\vbox to \y@p truein{        %% actual definition.
        \special{pos,inc=#2.fig}\vss }\hss }}}}\fi }    %%

%corrected section routine
\def\section#1{\par\ifnum\the\lastpenalty=30000\else
        \penalty-200\vskip\sectionskip\spacecheck\sectionminspace\fi
        \global\advance\sectionnumber by 1 
        \xdef\sectionlabel{\the\sectionstyle\the\sectionnumber}
        \wlog{\string\section\space\sectionlabel}
        \TableOfContentEntry s\sectionlabel{#1}
        \noindent {\caps\enspace\sectionlabel\quad #1}\par
        \nobreak\vskip\headskip\penalty 30000 }

%macros for figure insertion

\input epsf
\ifx\epsffile\undefined\message{(FIGURES WILL BE IGNORED)}
\def\insertfig#1#2{}% null macro
\else\message{(FIGURES WILL BE INCLUDED)}
\def\insertfig#1#2{{{\baselineskip=4pt
\midinsert\centerline{\epsfxsize=\hsize\epsffile{#2}}{{\centerline{#1}}}\
\medskip\endinsert}}}

\def\insertfigpage#1#2{{{\baselineskip=4pt
\pageinsert\centerline{\epsfysize=6.5in\epsffile{#1}}
 {{\bigskip{#2}}}\
\smallskip\endinsert}}}

\def\insertcomfigpage#1#2{{{\baselineskip=4pt
\pageinsert\centerline{\epsfysize=6.5in\epsffile{#1}}
 {{\smallskip{#2}}}\
\endinsert}}}

\def\insertccomfigpage#1#2{{{\baselineskip=4pt
\pageinsert\centerline{\epsfysize=6in\epsffile{#1}}
 {{\smallskip{#2}}}\
\endinsert}}}

\def\insertlongfigpage#1#2{{{\baselineskip=4pt
\pageinsert\centerline{\epsfysize=7in\epsffile{#1}}
 {{\medskip{#2}}}\
\endinsert}}}

\def\insertxfigpage#1#2{{{\baselineskip=4pt
\pageinsert\centerline{\epsfxsize=6in\epsffile{#1}}
 {{\bigskip{#2}}}\
\medskip\endinsert}}}

\def\insertsmallxfigpage#1#2{{{\baselineskip=4pt
\pageinsert\centerline{\epsfxsize=6in\epsffile{#1}}
 {{\bigskip{#2}}}\
\medskip\endinsert}}}

\def\bcap#1{{\bf Figure {#1}:\/}}

\fi

%%%%%%%%%%%%%%%%%%%%%%%%%
%  begin actual text    %
%%%%%%%%%%%%%%%%%%%%%%%%%

\CALT{2044}
\TITLE{Dynamical Simulation of Non-Abelian Cosmic Strings}
\AUTHOR{Patrick McGraw\footnote{\dag}{E-mail: pmcgraw@theory.caltech.edu}}
\CALTECH

\ABSTRACT{We describe a method for simulating the dynamics of an $S_3$ cosmic
string network.  We use a lattice Monte Carlo to generate initial conditions
for the network, which subsequently is allowed to relax continuously according to a simplified model of string dynamics. The dynamics incorporates some novel features which, to our knowledge, have not been studied in previous numerical simulations:  The existence of two types of string which may have different tensions, and the possibility that two non-commuting strings may intersect. 
Simulation of the non-commuting fluxes presents a computational challenge as it requires a rather complex gauge-fixing procedure.  The flux definitions change as strings
change their positions and orientations relative to each other and must be carefully updated as the network evolves.   The method is described here in some detail, with results to be presented elsewhere.}

\section{Introduction}
A generic feature of many spontaneously broken gauge theories is the existence
of topological solitons, including strings or flux tubes.  Many grand
unified models would predict the formation of such strings known as cosmic strings, during a cosmological phase transition.\Ref\Kibblecosmic{T.W.B.~Kibble, Phys. Rep. 67, 183 (1980).} 
  The many potential cosmological effects of cosmic 
strings\Ref\Cosmology{See, for example: A.~Vilenkin, Phys. Rep. 121, 263 (1985)} motivate one to wish to understand the subsequent evolution of such strings.  For example, since the gravitational effects of cosmic strings have occasionally been invoked as a possible source of density perturbations leading to galaxy or cluster formation, one is interested in knowing whether the strings are likely to decay very rapidly or persist long enough to seed density perturbations.  One is naturally also interested in the probability of observing strings in the present-day universe.  On the other hand, one might rule out certain phenomenological models if they seem to predict an abundance of strings inconsistent with present observations, much as the non-observation of monopoles has created difficulties for some models which predict a great abundance of monopoles.  

Some theoretical and numerical study \Ref\abeliansim{ T. Vachaspati and A.~Vilenkin, Phys. Rev. D 30, 2036 (1984) and references therein.  } has been devoted
 to the evolution of networks of Abelian strings such as those corresponding to the
Nielsen-Oleson \Ref\NielsenOleson{H.B. Nielsen and P. Oleson, Nucl. Phys. B 61, 45 (1973).} vortex solution of the Abelian Higgs model. It is generally believed that networks of this type are likely to form many closed loops which decay into smaller loops by the mechanism of intercommutation, resulting in a ``scaling'' distribution of loop sizes such that the number of loops within a cosmological horizon volume of the universe is roughly constant over time.  Vachaspati and Vilenkin \Ref\VV{T.~Vachaspati and A.~Vilenkin, Phys. Rev. D 35, 1131 (1986)} have performed a numerical simulation of a network of $Z_3$ strings, which have the novel feature that three strings may intersect in a vertex.  Their results suggested that such strings tend to form an infinite network of 
vertices connected by string segments.  The annihilation of these vertices leads to a similar scaling behavior, with the number of vertices and string segments
per horizon volume being roughly constant.  To my knowledge, however, very little is known about more complicated models in which the strings carry non-commutative magnetic fluxes and there is more than one type of string.  Such strings are known to exhibit a number of exotic types of interactions\Ref\various{See, for example, M. Alford, K-M Lee, J. March-Russell and J. Preskill, Nuc. Phys. B 384, 251 (1992)  and references therein.}.  Particularly significant is the fact that when two non-Abelian strings cross each other, they cannot generally intercommute, nor can they pass through one another without forming new vertices and becoming joined by a new segment of string.  Linked loops of string cannot usually become unlinked, and vice versa.  One might expect that this would inhibit the decay of a cosmic network by providing obstruction to the free removal of string segments.

In this paper, we describe a numerical simulation of a network of $S_3$ strings.
As in the $Z_3$ case, the strings form 3-string vertices.  In the spirit of reference [\VV ] we generate initial conditions from a lattice Monte Carlo simulation and then evolve the network according to highly simplified model of string dynamics which we hope captures the essential features of the system.  All string segments are assumed to be straight.  The vertices move under the influence of the string tensions acting on them.  We will see that a simulation of the non-commutative nature of the strings poses some rather difficult computational problems.  Only the method of the simulation is described here;  results will be presented elsewhere.\Ref\results{P.~McGraw, in preparation.}  Section {2} provides a brief summary of the properties of non-Abelian strings
 as they apply to this simulation.  We demonstrate some of the ambiguities inherent in the description of the state of a network of strings, and the necessity of a (gauge fixing) convention to resolve these ambiguities.  
Section {3} describes the particular $S_3$ model which we have chosen to
simulate.  In sections {4} and {5} we describe the gauge fixing convention  implemented by our program, and our method for making any needed comparisons of string fluxes in the simulated network.  Section {5} in particular discusses some of the subtleties that arise when periodic boundary conditions are used. Section {6} describes how we generate an initial condition for our simulation using a lattice Monte Carlo method.  The transcription of a lattice configuration into a  network that is fully specified according to the conventions of our simulation is itself a tricky task.  Finally, in section {7}, we describe the actual procedure for
simulating the time-evolution of the initial network, making reference to the conventions described in previous sections.

In this discussion, the strings will be considered as classical objects with
well-defined fluxes (after a gauge has been fixed).  We will not consider quantum-mechanical effects such as
Cheshire charge,\Ref\Cheshire{M. Alford, {\it et al.}, Phys Rev. Lett. 64, 1632 (1990); 65, 668 (E);  Nuc. Phys. B349, 414 (1991) } \/ even though the possibility exists that Cheshire charge might influence string dynamics.

\section{Vortices and strings in a non-Abelian discrete gauge theory}
Generically, topological defects of codimension 1 (vortices in two space
dimensions, strings in three)  occur when a gauge symmetry group $G$ is 
spontaneously broken to a subgroup $H$ such that there are non-contractible
closed loops in the vacuum manifold $G/H$.  This is formally expressed by saying
that $\pi_1(G/H)$ is nontrivial, where $\pi_1$ stands for the first homotopy
group.  This does in fact happen when $G$ is simply connected and $H$ is 
a discrete group.
Here, we review very briefly some of the properties of non-Abelian vortices and
strings which are important for the current simulation.  More details may be found in ref. \various.  
 
The breaking of an underlying $G$ gauge theory to a discrete group $H$ leaves no light propagating gauge fields:  At low energies in any simply connected region without defects,
the gauge field $A^a_\mu$ is pure gauge.  However, when string defects are present, the
region of true vacuum is not simply connected:  it is 
${\cal R} = {\cal M} - \{D\}$, where
$\{D\}$ is the union of all defect cores (regions of false vacuum)
 and ${\cal M}$ is the spatial manifold on which the defects exist.  Each string
gives rise to a class of noncontractible closed paths in ${\cal M} - \{D\}$ which encircle the string.  The flux enclosed by any closed loop $\Gamma$ is a group element defined as
a path-ordered exponential of the gauge field:  

$$ {\rm flux} = P \exp ( \oint_\Gamma A\cdot d\ell ).\eqn\pexp$$ 
For any $\Gamma$ within ${\cal R}$  this must be an element of $H$.  This is because the Higgs field is covariantly constant throughout ${\cal R}$ and so the
transformation that results from parallel transport around a loop must leave the
Higgs field invariant.  

The flux of a cosmic string is defined by the above exponential along a path which winds around the string.  In a non-Abelian theory, this definition of the
flux is not gauge invariant, and may depend on the point at which the path 
begins and ends.  A key observation, however, is that the flux through any loop
which does not enclose a string is necessarily trivial, and a corollary is that 
two closed loops which share the same beginning and ending point $x_0$, and can be continuously deformed into each other,   have the same
flux.  The relevant structure for the description of the system of defects is
the fundamental group or first homotopy group  $\pi_1({\cal M} - \{D\},x_0)$,
defined with respect to a basepoint $x_0$.  Each string is associated with a generator of the fundamental group.  Once $x_0$ has been (arbitrarily) chosen,  the fluxes of all closed paths (and of all strings) are specified by a homomorphism
from  $\pi_1({\cal M} - \{D\},x_0)$ into $H$.  The only remaining gauge freedom
is a global one.  However, there is a considerable amount of ambiguity in what we mean by ``the flux'' of one particular string:  an arbitrariness in how exactly
 the set of generators is chosen for the homotopy group.  In figure \FIG\Genambig ~\Genambig \/, 
for example, there are two loops, both beginning and ending at $x_0$, both enclosing the same string without enclosing any others, which are nonetheless
representatives of different homotopy classes (and consequently may be associated with different fluxes):  Another intervening string prevents one
path from being continuosly deformed to the other.  The fluxes associated with
the two different paths may differ through conjugation by the flux of the other
string.  In an Abelian theory, conjugation is trivial;  not so in a non-Abelian one.  It follows that fluxes cannot meaningfully be compared (say, to determine
if they are the same) if the paths used to define those fluxes pass on opposite
sides of some other string.  Comparisons must be made using ``nearby'' paths.

\insertxfigpage{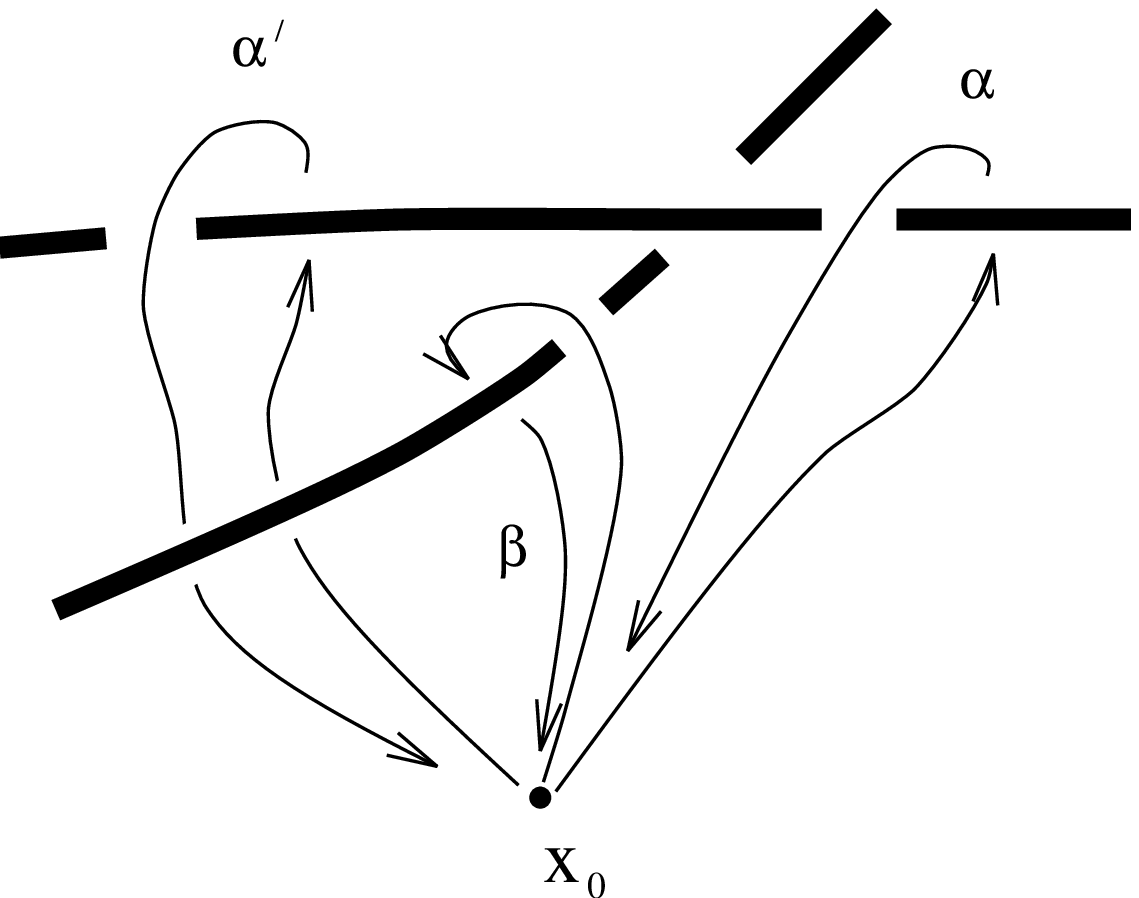}{\bcap\Genambig \/  The paths $\alpha$ and 
$\alpha '$  both enclose the same string and no other strings,  but they cannot
be continuously deformed into each other without crossing another string.  Thus,
they represent different elements of the fundamental group
$\pi_1({\cal M} - \{D\},x_0)$, and so the fluxes associated with them may be 
different.  Specifically, the homotopy classes of $\alpha$ and $\alpha '$ are related through conjugation by another generator: 
$\alpha' \sim \beta\alpha\beta^{-1}$.  (We follow the usual convention of
composing paths from right to left:  $\beta\alpha\beta^{-1}$ means the path formed by traversing first the reverse of $\beta$, then $\alpha$, then $beta$.
The relation $\sim$ represents homotopy equivalence.)
The associated fluxes are analogously related:
a nontrivial relation if the fluxes don't commute. }

In the sequel, the establishment of a conventional set of generators for
 $\pi_1({\cal M} - \{D\},x_0)$ and determining their associated $H$ elements is often
referred to somewhat loosely as ``fixing a gauge.''  As a notational convention, we will usually label the representatives of particular homotopy classes with Greek letters, while using Roman letters for the associated $H$ elements.

The dependence of the definition of a string's flux on other strings between it 
and the basepoint, as described above, has important consequences for the behavior of a network of non-Abelian strings.  One of these is the occurrence of
so-called ``holonomy interactions'' or exotic exchange interactions.  These are a consequence of the fact that as strings move through space, the paths which are
most convenient for describing their fluxes may be required to cross other strings.  Strings may thus change their flux quantum numbers simply by changing
their positions with respect to other strings.  An example is shown in figure
\FIG\interchange ~\interchange \/,  where two strings rotate around each other
and thus change their fluxes.  Since non-Abelian strings with conjugate fluxes
may be transformed into one another through holonomy interactions or by 
gauge transformations,  all strings with fluxes in a given conjugacy class
must be degenerate with each other in terms of their tensions.  Different
conjugacy classes of $H$, however, may be associated with different types of strings.  

\insertxfigpage{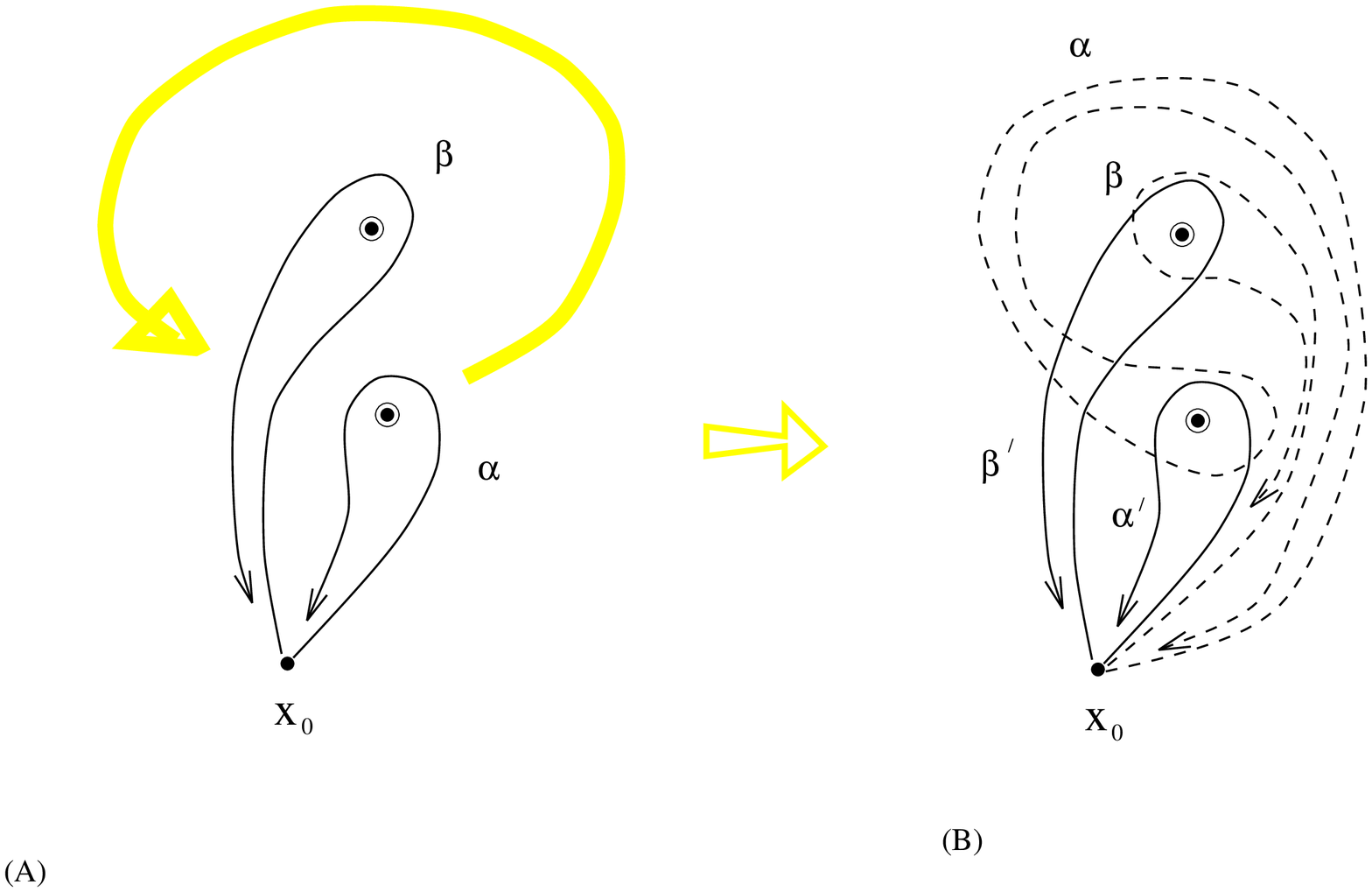}{\bcap\interchange \/  A holonomy interaction:
we show here a planar cross-section of a process in which two strings wind around each other.  If two strings (or rather their intersections with the plane) are initially in the positions shown in (A), and one of them winds completely around the other, then the path $\alpha$ shown in (A) can be deformed to the dotted path $\alpha$ in (B) without crossing any strings.  However, the path shown as $\alpha'$ is the same path {\it in space} as the old path $\alpha$, so it makes sense to redefine the flux of the string according to
this new path, which is actually homotopically equivalent to 
$\beta\alpha\beta^{-1}$.  Therefore the new flux of one string is conjugated by the flux of the other:  if $a,b,a',$ and $b'$ are the fluxes associated with
$\alpha, \beta, \alpha'$ and $\beta'$ respectively, then $a'=bab^{-1}$.  In this figure, the flux of the other string is also affected:  
$\beta' \sim \alpha'\beta(\alpha')^{-1}$ and hence $b'=a'b(a')^{-1}=(ba)b(ba)^{-1}$.  Note that the product of the fluxes is conserved: $b'a'=ba$. This is to be expected as $\beta\alpha$ can be deformed
to a loop that completely encircles the pair and need not be affected by their
relative motion. }

Another consequence of the non-Abelian interactions is that two strings with
noncommuting flux cannot intercommute, nor can they pass through each other 
unaffected;  to do so would violate flux conservation.  As illustrated in figure \FIG\generalNCI  ~\generalNCI \/, noncommuting strings can only pass
through each other if a new string segment is formed, linking the two strings
to each other and carrying a flux which is the commutator of the fluxes of the two original strings.  We will be especially interested in the consequences of
this entanglement process for the evolution of a string network:  it might 
impede the collapse of the network.

\insertfigpage{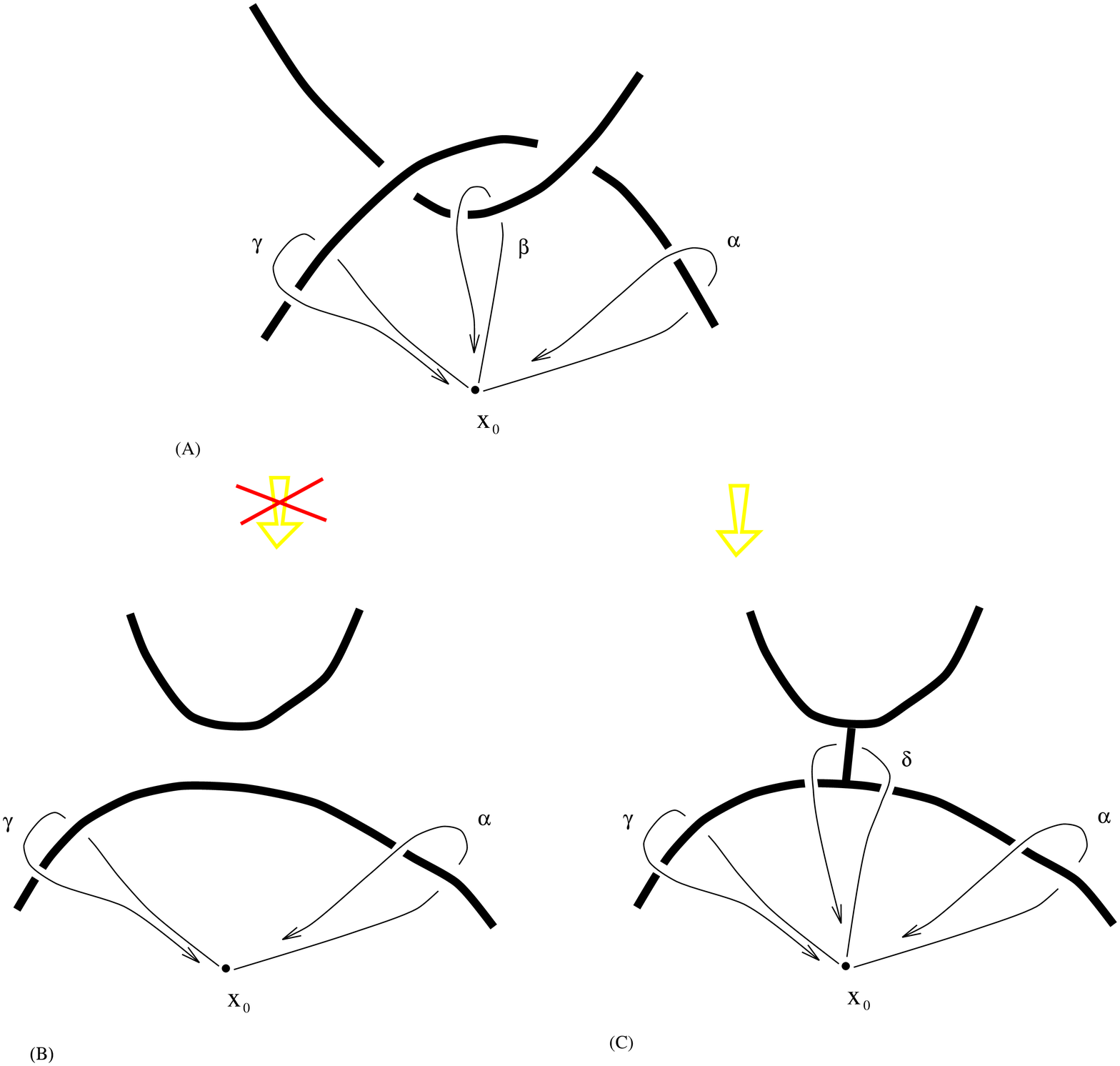}{\bcap\generalNCI \/ Attempt to pass two strings through each other. In (A) the flux of one string may be defined by either of 
the paths $\alpha$ or $\gamma$, and that of the other string by $\beta$.  Let
the  fluxes associated with $\alpha$, $\beta$ and $\gamma$ be $a,b,$ and $c$
respectively.  In this case, $c=bab^{-1}$.  In general, $c\neq a$. Now, if we attempt to pass the strings through each other,  no strings need cross the paths $\alpha$ and $\gamma$,  so the associated fluxes will not change.  But if the strings were to pass through each other freely, as in (B),  $\alpha$ and $\gamma$  would be continuosly deformable into each other.  This is impossible
if they have different fluxes.  In order to conserve flux,  the string must branch somewhere and be connected to the other by a new string whose flux
as defined by path $\delta$ in (C) is 
$ca^{-1}=bab^{-1}a^{-1}$.}

\section{Our Model: $S_3$ Strings}
We consider here a model with unbroken gauge group $H=S_3$, the permutation group
on three objects.  The spectrum of this model will include strings  whose fluxes
are elements of $S_3$.  $S_3$ has six elements in all.  The identity corresponds to the trivial permutation.  There are three odd permutations (two-cycles or transpositions) each leaving one of the three elements invariant and interchanging the other two.  We may denote these, for convenience, by:
$t_1=\{(123)->(132)\}, t_2=\{(123)->(321)\}, t_3=\{(123)->(213)\}$.  In this notation, $t_i$ is the two-cycle which leaves the $i$-th element in the same position.  The two
non-trivial even permutations are the three-cycles, or cyclic permutations,
which we denote here by  $s_+=\{(123)->(312)\}, s_-=\{(123)->(231)\}$.  In the more conventional cycle notation\Ref\Herstein{I.N.~Herstein, Abstract Algebra. Macmillan, New York 1986}, we 
have $t_1=(23), t_2=(13), t_3=(12), s_+=(123), s_-=(132)$.  

The 2-cycles form one of the two non-trivial conjugacy classes, and the 3-cycles form another.
Thus our model supports two types of strings, which we shall refer to as $t$-strings and $s$-strings, respectively.  The three-cycles generate a $Z_3$
subgroup, so that any system containing only $s$-strings will behave identically
to the $Z_3$ string system previously studied in [ \VV ].  There can be junctions where three $s$-strings join.  Another type of junction is one where
two 2-cycle (or $t$) strings merge to form a 3-cycle ($s$) string.  Since each two-cycle is equal to its inverse, oppositely oriented $t$-strings are topologically equivalent.  $s$-strings, on the other hand, possess a natural orientation:  The flux through a path encircling it with one orientation is $s+$, while it is $s-$ for the opposite orientation.  In subsequent figures, $s$-strings will often be denoted by oriented lines, with the string carrying
flux $s+$ in the direction of the arrow, while t-strings have no arrow.  Figure
\FIG\junctions ~\junctions \/ shows the two types of junctions in our model.

\insertfigpage{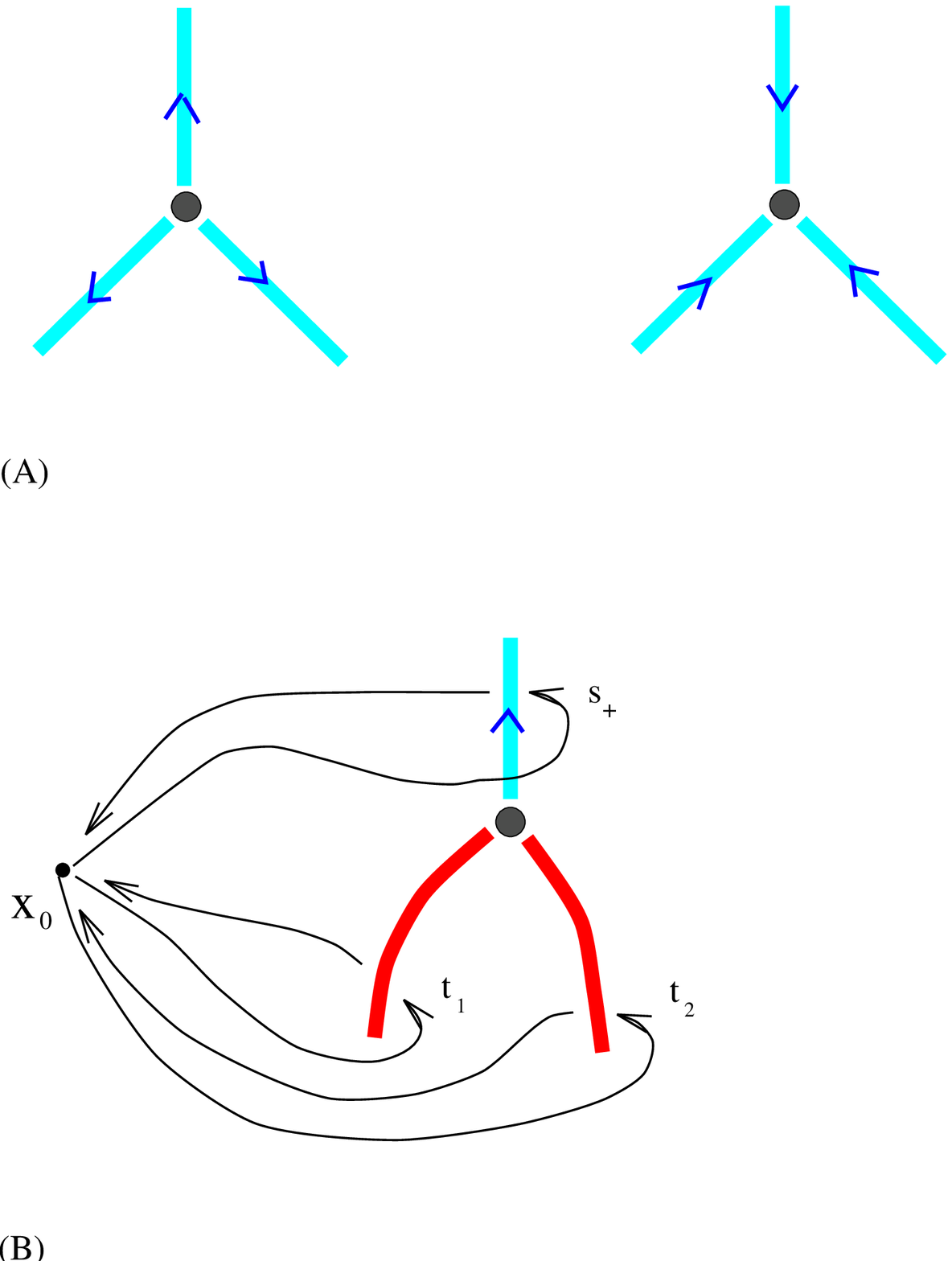}{\bcap\junctions \/ String junctions in the $S_3$ model.  (A) Two possible $sss$ junctions:  three strings with the same flux, $s_+$ or $s_-$, emanate from the node. (Or two $s_+$ strings merge into a single $s_-$, etc.)  (B) One of the class of $stt$ junctions:  Two $t$-strings merge into an $s$-string.  Fluxes are defined with respect to $x_0$ by the paths shown.  Here, as in many subsequent figures, an $s$-string is drawn as an oriented line.  The string carries flux $s_+$ in the direction of the arrow: i.e., a counterclockwise path around the arrow encloses flux $s_+$}

The system we propose to simulate consists of $s$ and $t$ strings joined together at two types of vertices ($sss$ and $stt$) which we shall take to be two types of monopoles.  The tensions of all $t$-strings are the same, as are
the tensions of all $s$-strings.  The ratio of these two tensions will presumably depend on the details of the Higgs mechanism which produces the strings, and we will take it to be an adjustable parameter of the model.

In simulating the dynamical evolution, the strings will be approximated as straight segments between junctions.  This amounts to averaging over any oscillations of the strings.  We will also assume, following ref. \VV, that the string junctions
undergo damped motion under the influence of string tensions.  This assumption may be a crude approximation to the dynamics of any real network.  It becomes realistic if the monopoles are imagined as carrying some unconfined magnetic flux. For example, the actual symmetry-breaking pattern may
be locally $G->S_3 \times (SU(3)\times SU(2)\times U(1)_{EM})$, with some discrete factor divided out so that the monopoles at string junctions may carry electromagnetic $U(1)$ charge.  Their magnetic charges should then result in radiation damping.  Such a pattern has been demonstrated in a model where 
topological $Z_n$ strings become attached to monopoles which also carry other charges.\Ref\coloredmonopoles{M. Hindmarsh and T.W.B. Kibble, Phys. Rev. D 55, 2398 (1985)}  It is possible in principle for $S_3$ strings to join at monopoles, although it may require a more complicated model.  For example,  consider
the monopoles that form when an $SU(5)$ group is broken in the familiar way to $SU(3) \times SU(2) \times U(1)/Z_6$.    This transition is
known to yield stable monopoles with $SU(3), SU(2),$ and $U(1)$ flux.\Ref\monopoles{M.Daniel, G.Lazarides, and Q. Shafi, Nuc. Phys. B170, 156
(1979)} \Ref\SUfivebreak{J. Preskill, Ann. Rev. Nucl. Part. Sci., 34,  461 (1984)}  We could imagine a second symmetry-breaking stage in which the 
$SU(3)/Z_3$ factor is broken down to $S_3$ in such a way that the resulting srings also carry nontrivial flux in the $Z_2$ center of $SU(2)$.  Whenever
three such strings join, the resulting net  $Z_2$ flux can unwind through a monopole, which 
has both $SU(2)$ and $U(1)$ flux.   

The masses and unconfined charges of the two types of monopoles may be model-dependent parameters relevant to the network's evolution.

\section{Gauge Fixing Conventions}
The present simulation requires that we choose some convention by which to
define the fluxes of all strings in the network, and keep track of the 
evolution of those fluxes as the strings and nodes move. 
 
In our algorithm,  the strings and nodes exist inside a rectangular volume with opposite sides identified: a 3-torus.  The subtleties associated with the  periodic boundary conditions will be discussed later: for now we simply consider a network inside a rectangular volume with boundaries.  We choose a cubic volume
with one corner at $(0,0,0)$ and the far corner at $(L,L,L)$. We choose a basepoint at the center of our simulation 
volume, $(L/2,L/2,L/2)$.  Let each node be associated with a straight line segment (a ``tail'')
along the direction ${\vec{BN}}$ from the basepoint to the node's location.  Then let the flux
of each outgoing string be defined with respect to a path which runs
outward along this tail to a point which is taken to be vanishingly close
to the the node.  The path then encircles the string in a counterclockwise direction and returns to the basepoint along the node's tail. This is 
illustrated in figure \FIG\tail ~ \tail.  This will be our convention for defining the fluxes of the strings which join at a given junction.

\insertfigpage{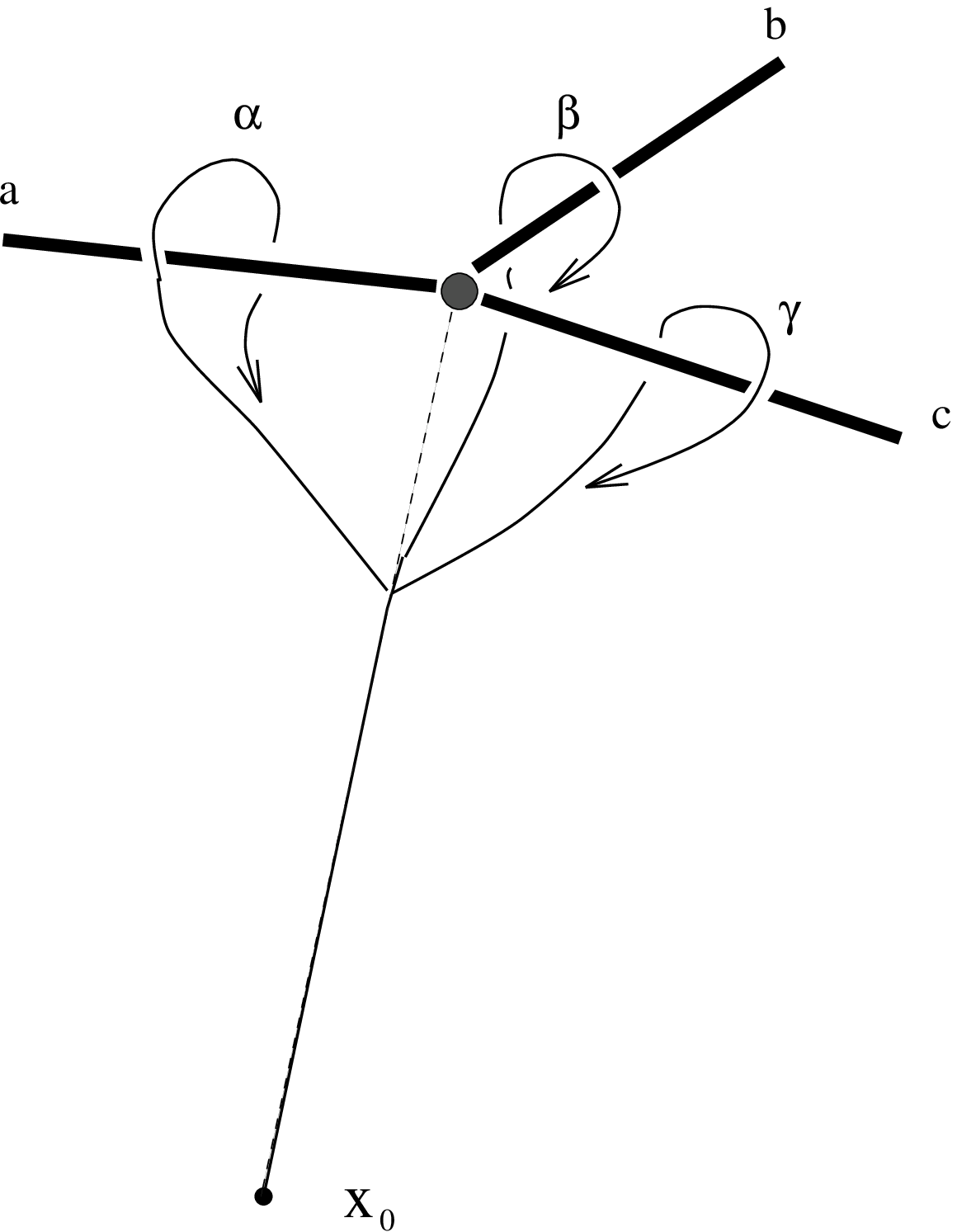}{\bcap\tail \/ Conventions for measuring the fluxes of the three strings emanating from a
node.  Each string's flux is defined as the flux through a path which leaves the basepoint $x_0$ along a straight line toward the node, then encircles the string in
a counterclockwise direction as seen from the far end of the string and returns
to the basepoint.
} 

 As illustrated in figure \FIG\product ~\product, flux conservation requires that the product of all three fluxes emanating from a node be trivial when the fluxes
are multiplied in a clockwise order with respect to the direction ${\vec{BN}}$.
i.e., if the strings in clockwise order are $a$, $b$, and $c$, then
$$cba = e.\eqn\fluxcons$$ 
  In our algorithm, a record is maintained of the geometry of each node:  the strings carry labels indicating the appropriate clockwise orientation.

It is possible for a pair of nodes to be connected by more than one string, as shown in figure \FIG\doubleone ~\doubleone.  In this case, the two (or more)
segments are treated as collinear, and the order is therefore ill-defined.  In such a case we allow the order to be arbitrary, but the fluxes of the two strings must be defined in such a way consistent with that order, such that the product of all three fluxes is as usual trivial.  The ordering must also be compatible between the two nodes which the segments join, so that the flux of a given segment is consistent at its two ends. (The consistency of segments from one end to another will be discussed below.)

\insertfigpage{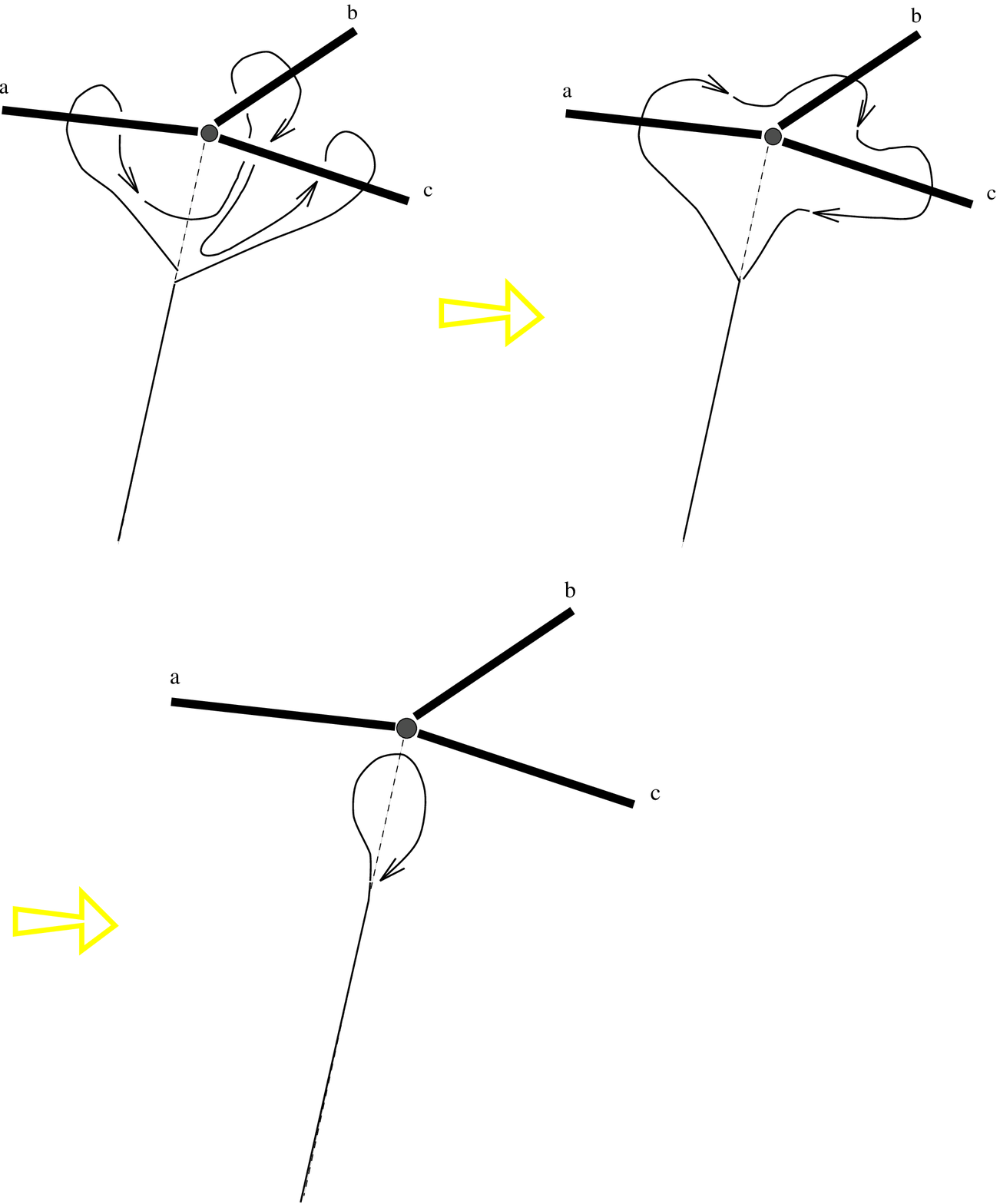}{{\bf Figure \product:}\/  The composition $\gamma\beta\alpha$  of all three paths can be continuously
deformed to a point.  Therefore $cba$, the product of all three fluxes taken in a
clockwise direction as seen from above the node, must be trivial.}

\insertxfigpage{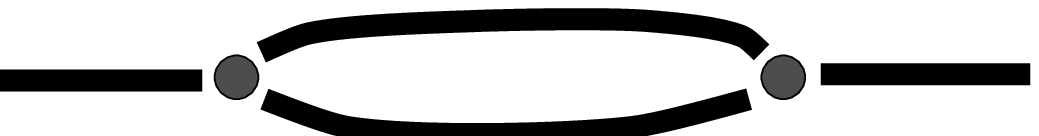}{\bcap\doubleone  \/  Doubly linked nodes.}

  The collection of standard paths defined above represents a set of generators for $\pi_1({\cal M}-\{D\})$.    The flux state of a network of strings is fully
specified when we know the fluxes enclosed by all of these standard paths. The condition \fluxcons \/ supplies one set of 
relations among these generators.  For each string segment, there is also a relation involving the fluxes defined at its two endpoints, as discussed below.

\subsection{``Sliding'' flux from the endpoint}

By the conventions above, the flux of each string is defined at its two
endpoints.  But for the purposes of this simulation it will be necessary to make comparisons of the fluxes of strings at arbitrary points along their lengths.  For example, if two strings cross each other, it will be necessary to determine whether or not their fluxes commute.  A meaningful comparison of the fluxes of
nearby string segments can be obtained only if the paths used to define those two fluxes remain close to each other everywhere except in the immediate vicinity of the strings to be compared.   In particular, the ``tails'' of the paths must not pass on opposite sides of any string, because such paths would give different flux measurements for the same string.     It is possible to define the flux of a string at an arbitrary point
along its length by sliding the standard path to the one which encircles the 
string at the point we wish to measure, as illustrated in figure \FIG\basicslide ~\basicslide.  If another string with flux $b$ pierces the triangle
which is swept out by the sliding path, then the flux at the new position is conjugated by b.  If multiple strings occur, then the new flux $a'$ is given by $faf^{-1}$,  where $f$, the total flux inside the triangle, is defined as the product of the fluxes of all enclosed strings, taken in order of increasing angle from the initial ray $\vec{BP}$.  The flux of each other string at the point where it pierces
the triangle must in turn be defined by a similar sliding procedure from one of
its ends.  This procedure, applied recursively, can thus define the flux of any
string at an arbitrary point $P$ along its length, as measured by a path which follows a straight line from $x_0$ towards $P$ and encircles the string near 
$P$.  If one slides the path all the way to the far end of the string, the resulting value of the flux must be consistent with the value
measured by the standard path at the other end.  This specifies an additional set of relations among the generators of  $\pi_1({\cal M}-\{D\})$ \/ and furnishes one way of testing for errors in the simulation.

\insertfigpage{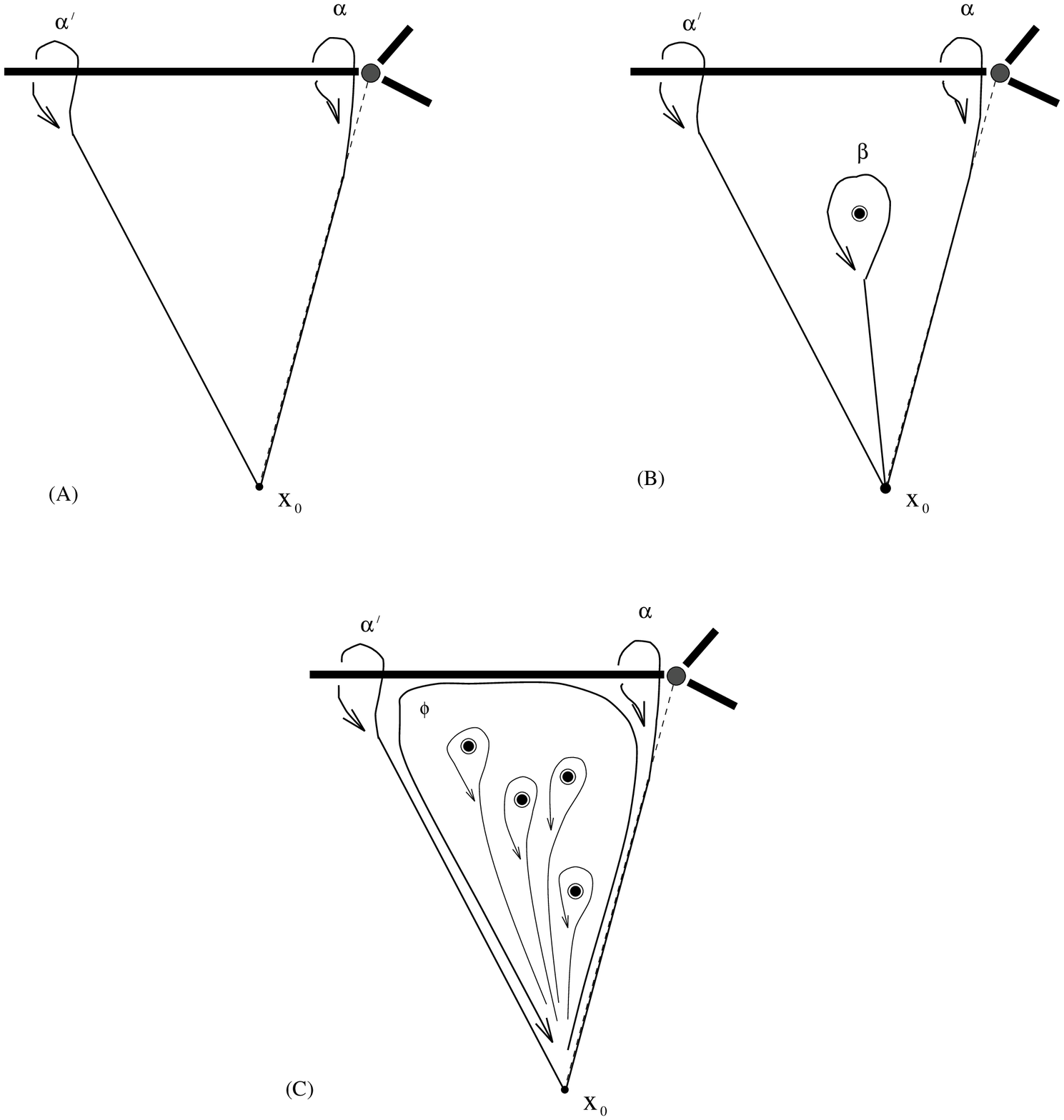}{\bcap\basicslide \/ 
When the flux $a$ of a string has been defined according to a path which encircles
it near one end,  the flux $a'$ of that string at another point along its length  can be defined by ``sliding'' the standard path $\alpha$ to $\alpha'$ as shown.
If no other strings pierce the triangle which is swept out, then this merely
represents a continuous deformation of $\alpha$, and thus $a'=a$\/ (fig. \basicslide A).  However, if the triangle is pierced by string with flux $b$ as measured by path $\beta$,
then the flux is conjugated by $b$:  $a'=bab^{-1}$\/ (\basicslide B).  More generally, if the 
triangle (or the oriented path $\phi$ shown in \basicslide C) encloses flux $f$, then $a$
is conjugated by the total flux $f$, i.e., $a'=faf^{-1}$. The total flux is given by the product of individual string fluxes, taken in order of increasing
angle from the initial tail. (This can be seen by deforming a product of loops
to a single loop enclosing all strings.)} 

\subsection{Holonomy Interactions}

As the network evolves dynamically and nodes change their position, it is
possible for the fluxes defined by these conventions to change by several
different mechanisms.  First, as a node moves, its tail may be dragged across
another string segment.  Conversely, a string segment may be dragged across
the node's tail by the motion of other nodes.  In both cases, the fluxes of
all strings at the node must be conjugated by the flux which is crossed, as
shown in figure \FIG\tailcrossing ~\tailcrossing.  In addition, the geometry of the strings
at a given junction may change, resulting in holonomy interactions among the 
three strings joined at that node.  Such a process is shown in figure
\FIG\geochange ~\geochange :  the motion of string $a$ causes its standard flux to change, 
and also changes the clockwise ordering of the strings $a$, $b$,and $c$. 

\insertfigpage{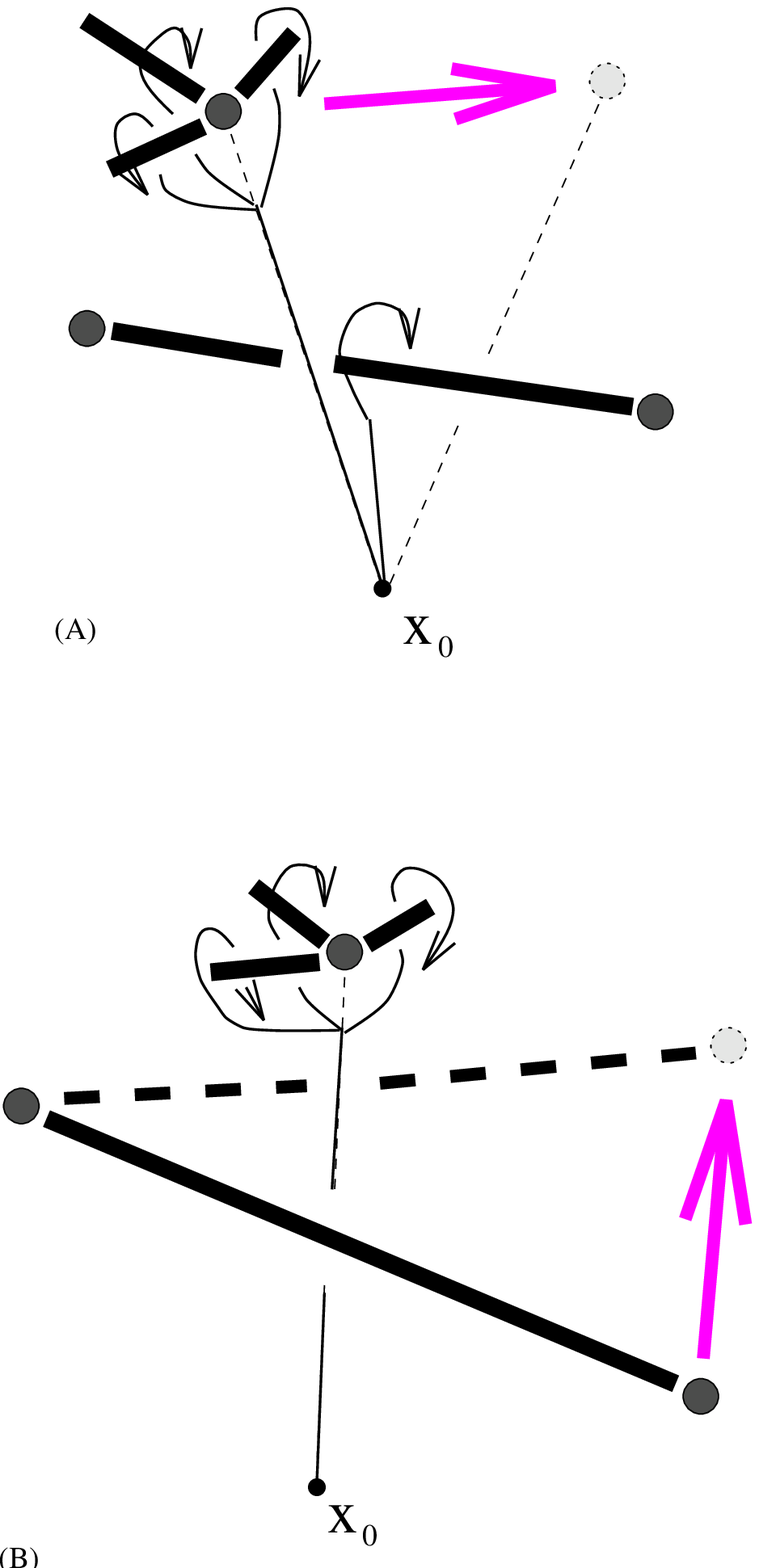}{\bcap\tailcrossing \/ Crossing of a node's tail
by a string.  This can happen either when the moving node drags its tail across
the string (A), or when the string is dragged across the tail due to the motion
of another node (B).  In both cases, the fluxes of all strings attached to the 
node whose tail is crossed must be conjugated by the flux of the crossing string.} 

\insertxfigpage{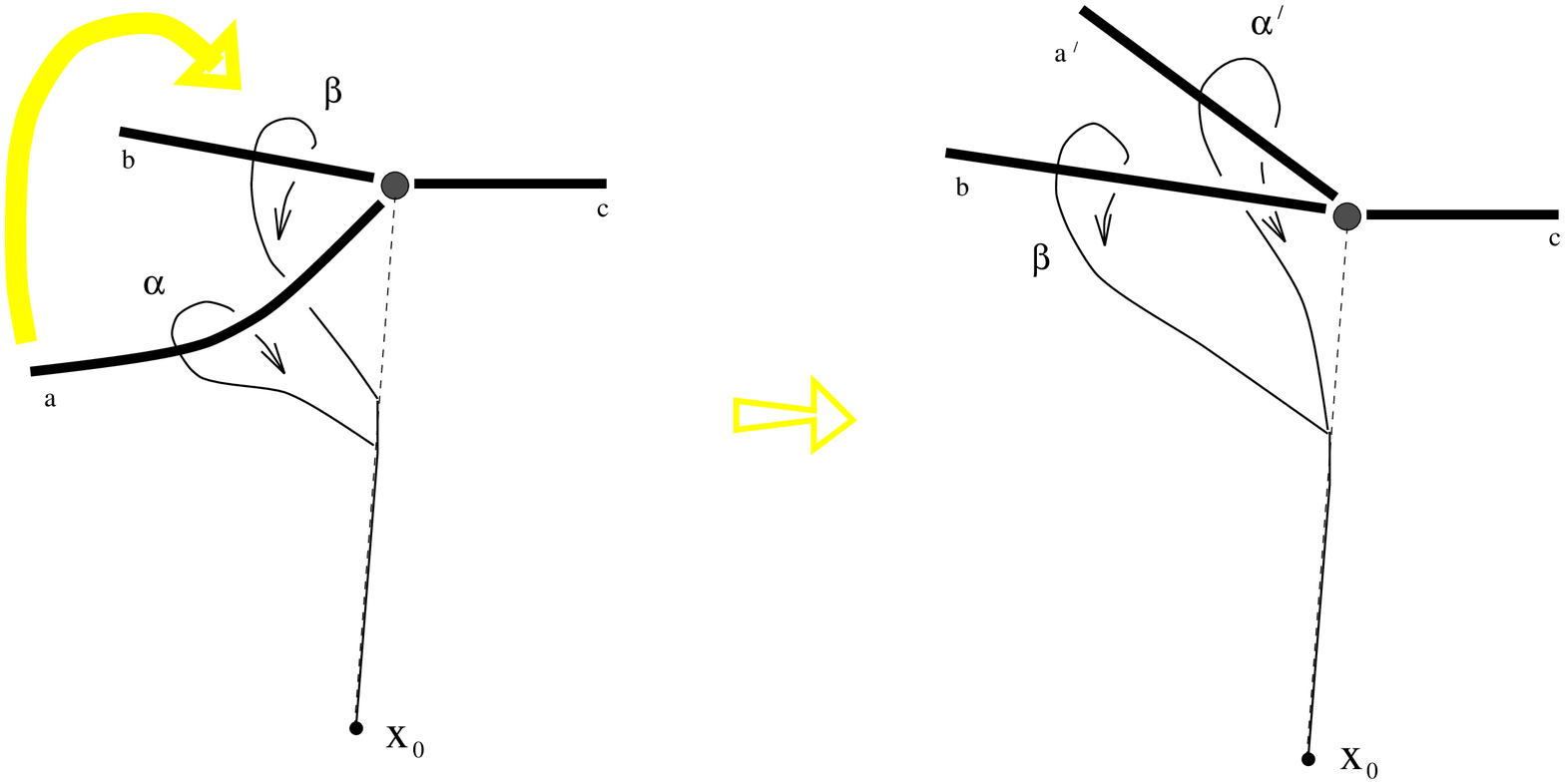}{\bcap\geochange \/ Example of a holonomy interaction between strings attached to the same node.  When the string carrying flux $a$ is lifted over the other string carrying flux $b$, its flux must be
redefined as $bab^{-1}$, and the conventional clockwise order of the three strings changes, with $a$ and $b$ exchanging places.  The flux conservation condition is maintained:  if $cba=e$ originally, then also $ca'b=e$.}

\section{Periodic Boundary Conditions}
In order to maintain isotropy everywhere in the simulation volume, we use periodic boundary conditions.  The cube face $x=0$ is identified with $x=L$,
and similarly for $y$ and $z$.  This identification has two consequences.
The first is that there are three additional classes of noncontractible closed loops starting and ending at the basepoint: namely those which wrap around one
of the boundaries before returning.  It is possible for these loops to be associated with nontrivial flux, and a complete specification of the state of our network requires that we choose representatives of these classes and maintain a record of the associated fluxes.  The second consequence is that nodes may
move freely across the cube boundaries, and string segments may wrap
around from one side of the cube to another.

As representatives of the three ``wraparound'' classes, we
choose straight-line paths which we will refer to as $\Gamma_x,\/\Gamma_y$, and 
$\Gamma_z$.  $\Gamma_x$, for instance, leaves the basepoint along the $+\hat{x}$ direction, wraps around the boundary from $x=L$ to $x=0$, and then returns to the basepoint from
the $-\hat{x}$ side.  

 A record of the fluxes $C_x, C_y,$ and $C_z$ associated with these three paths,
combined with the record of all string fluxes as defined earlier, specifies the
state of the string network on $T_3$.  In order to complete the description, however, there is one ambigutity to be resolved.  The fluxes of strings, as 
described above are measured along paths which follow a ``tail'' from the basepoint to the point where the flux is to be measured, then encircle the flux 
and return to the basepoint.  On $T_3$, however, a segment from $x_0$ to an 
arbitrary point $x$ is not unique, as shown in figure \FIG\notunique ~\notunique.  The 
two points may be connected by a line segment which does not wrap around, or by
one which does.  (In fact, there is an infinite set of possibilities.)  We may 
choose to define all tails in such a way that none of them cross the boundary,
but it is still necessary to make flux comparisons across the boundary, and
it is necessary to redefine fluxes when a node moves across the boundary.  In
short, is necessary to relate the alternative definition of a string's flux
which are related to one another by wrapping.  A prescription for doing so is
shown in figure \FIG\wrapflux ~\wrapflux:  The transformation from one description to
the other requires us to know the flux associated with one of the $\Gamma_i$.

Much as holonomy interactions may change the measured fluxes of strings,  similar effects compel a redefinition of $\Gamma_i$ when a string segment is
dragged across the standard path.  As illustrated in figure \FIG\gammacrossing ~\gammacrossing,  the new flux value $C'_i$ will be a product of the old value
$C_i$ with the flux of the string which crosses it.  (The order in which the two
are multiplied depends on the handedness of the crossing and on which side of the
basepoint it occurs.)

\insertxfigpage{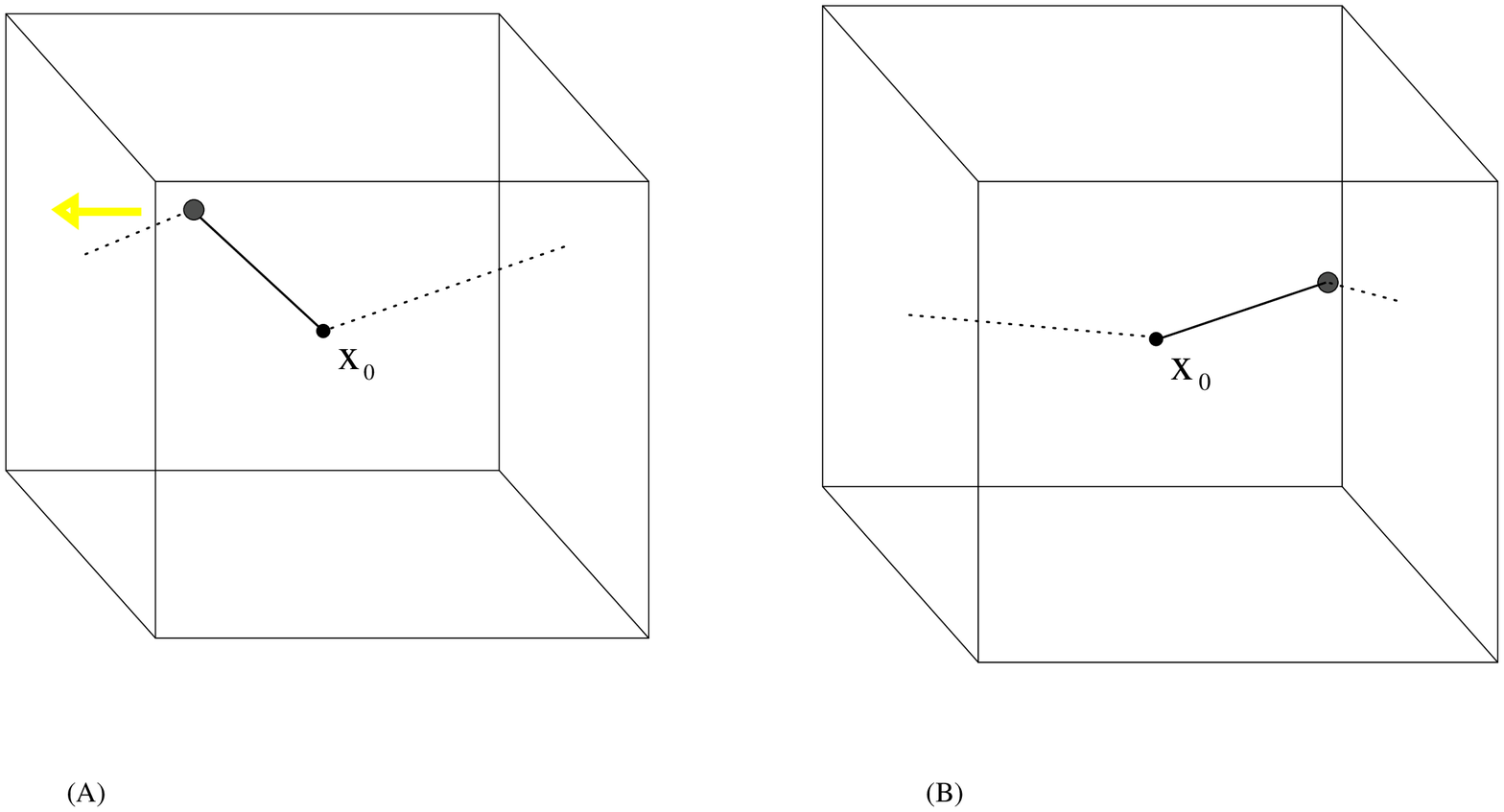}{\bcap\notunique\/  Under periodic boundary 
conditions, there is more than one straight line segment from the basepoint
to a given node.  In (A), we show one segment which extends to the left of 
$x_0$ and ends at a node (solid line), and another which extends to the right, wraps around
the cube boundary, and ends at the same node (dotted line).  By convention, we will choose to
describe nodes using the shortest possible segment as a ``tail'' (shown here as the 
solid line).  If the node moves to the left as shown by the arrow, and wraps around the cube
boundary to reach the final state shown in (B), then the shortest segment will no longer be the one extending to the left from $x_0$, but instead it will be the one extending to the right.  The new shortest segment is shown as the solid line
in (B).
Because of this change in the choice of ``tail,''  it is necessary to know how to transform a flux from one description
to the other.}

\insertccomfigpage{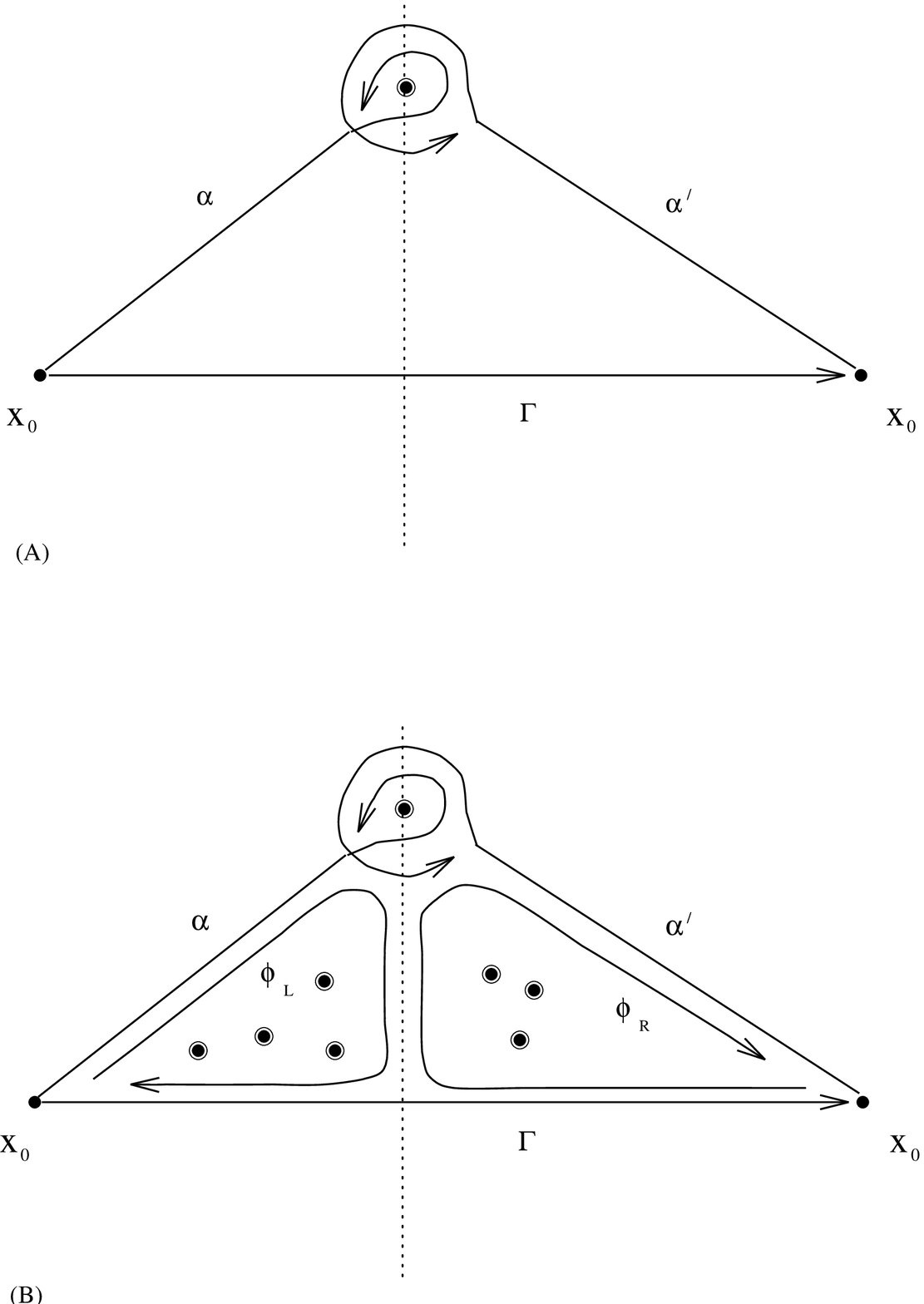}{\baselineskip=7pt \bcap\wrapflux \/ Transformation from one description of a flux to another at the boundary.  Here a string is shown intersecting the plane of the page precisely where it intersects the boundary of the cubic simulation volume (dotted line).  Under periodic boundary conditions, the two points labeled $x_0$ are identified. The flux of the string may be described in terms of a path whose tail extends to the right of $x_0$ ($\alpha$) or to the left ($\alpha'$).  If no other strings are present, then $\alpha$ is homotopically
equivalent to $\Gamma^{-1}\alpha'\Gamma$.  In the more general situation shown in (B), $\alpha\sim (\phi_L \Gamma \phi_R)^{-1}\alpha' (\phi_L \Gamma \phi_R)$, and so the two descriptions of the flux are related through conjugation by $f_L C f_R$,  where $C$ is the flux associated with the path $\Gamma$ and $f_L$ and $f_R$ are the overall fluxes enclosed by $\phi_L$ and $\phi_R$, respectively.  The latter can be defined in terms of paths lying entirely on one side or the other of the boundary.}

\insertcomfigpage{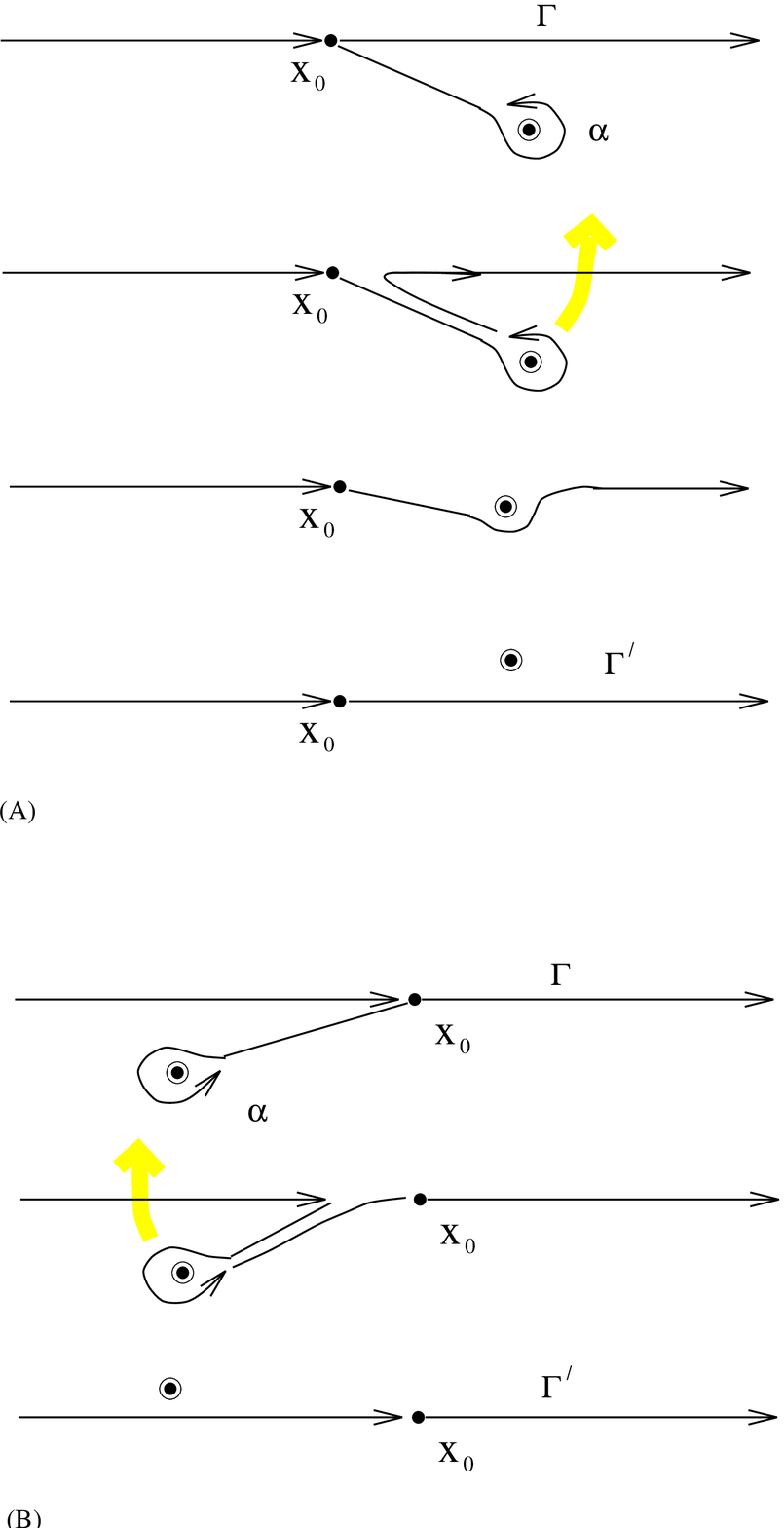}{\bcap\gammacrossing \/ Interaction between a string and one of the large loops of the 3-torus.  As a string with flux $a$ as defined by the path $\alpha$ crosses to the right of $x_0$ as shown in (A), the path $\alpha\Gamma$, where $\Gamma$ is a straight-line path which wraps around the 3-torus, can be continuously deformed to the new straight-line path 
$\Gamma'$.  Thus the flux $C$ associated with $\Gamma$ must be multiplied from the right by the string's flux, $C'=Ca$.  If the string crosses to the left as in (B), $\alpha\Gamma$ is deformed to $\Gamma'$ and so the multiplication is from the other side: $C'=aC$.}

The existence of the three closed paths which wrap around the boundary, and of
holonomy interactions between the fluxes of these paths and those of strings,
make this system a higher-dimensional analogue of the system of vortices on a Riemann surface which was previously studied by Lee.\Ref\Lee{K.M.~Lee, Phys. Rev. D 49, 2030 (1994)}

\section{Initial Conditions}
In order to perform our dynamical simulation, we must start with a randomly generated initial configuration.  The generation of initial conditions should to some extent be a model of the symmetry-breaking transition which produces the strings.  Following the lead of Vachaspati and Vilenkin's $Z_3$  simulation, we use a lattice to generate an initial string
network.  The lattice spacing is to be identified in our minds with the correlation length of the Higgs field which acquires a vacuum expectation value in order to break a continuous gauge group to a discrete subgroup H.  The Higgs
VEV is thus uncorrelated over distances longer than a lattice spacing, and at each site of the lattice, it takes a random value within the vacuum manifold.  With a suitable interpolation along the length of each link, any plaquette of the lattice is mapped to some closed loop on the vacuum manifold.  If this path is one of the non-contractable loops, then a string must pierce this plaquette.  Each link of the plaquette is associated with an element of $G$ which relates the Higgs values at the two ends of the link.  The product of these elements must lie within the unbroken group H, and can be taken as the flux of the string which pierces that plaquette. Strings which pierce the faces of a given unit cube must be joined together inside the cube in some appropriate way.  If only two faces of the cube are non-trivial, then we interpret this as a single string segment passing through the cube.  If three faces are non-trivial, and our model is one in which three-string vertices occur, then we conclude that there is a single vertex inside the cube.  Cubes pierced by more than three ends require a more complicated arrangement of nodes and strings inside the cube, and there may be more than one self-consistent way to join the string ends.  In the
following subsections, we will first describe methods for simulating on a lattice the string-forming phase transition, and then how to translate the results into a string network which can serve as the initial condition for our 
dynamical evolution algorithm.

\subsection{Discrete Higgs Simulation}

Generating random points on an arbitrary continuous vacuum manifold and then interpolating suitably between them can be a difficult proposition.
A useful technique is to approximate the broken group $G$ by a discrete group ${\cal G} \subset G$ which contains the unbroken group $H$ as a subgroup.\REF\DH{T.~Vachaspati and A.~Vilenkin, Phys. Rev. D 30, 2036 (1984).
T.W.B.~Kibble, Phys. Lett. B 166, 311 (1986)} \REF\Zinitial{M.~Aryal, A.E.~Everett, A.~Vilenkin and T.~Vachaspati, Phys. Rev. D 34, 434 (1986)}
\refmark{\DH,\Zinitial}
    Instead of a continuum of values, the Higgs field takes values in a discrete coset space ${\cal G}/H$.  Then with each link of the lattice there is associated an element of ${\cal G}$ which transforms the Higgs field value at one end of the link to the value at the other.  The element relating one coset
to another is not unique;  the possible elements themselves form a coset.  The
convention in this discrete Higgs method is to choose the ``smallest'' possible element for each link variable.  ``Small'' is defined with reference to a metric
on the continuous group $G$:  if all elements are written in the form 
$g=\exp (i \alpha T)$ where $T$ is a normalized element of the Lie algebra of
$G$, then the smallest element is the one with the smallest number $\alpha$.
In this way the Higgs field is effectively interpolated in the smoothest possible way between lattice points.   A suitable gauge transformation can be performed so that all Higgs field values lie in the same coset, and all link variables lie within $H$, allowing all subsequent computations to be performed 
in terms of only $H$ link variables.  This was the technique used in previous studies of $Z_3$ networks, \refmark\Zinitial \/  and we will use it here, taking ${\cal G}$ to be
one of the discrete subgroups of $SU(3)$.  
For example, we may use a 24-element ``dihedral-like'' subgroup of $SU(3)$ known as $\Delta(24)$. \Ref\SUsub{W.M.~Fairbairn,T.~Fulton, and W.~H.~Klink, J.Math.Phys. 5, 1038 (1964)} This group is isomorphic to $S_4$, the permutation group on four elements, and is the smallest discrete subgroup of $SU(3)$ which contains $S_3$ as a proper subgroup.

An easier way to generate a random network of strings is to use an infinite temperature lattice gauge theory:  simply assign a random element of the unbroken group $H$ to each link of the lattice, and evaluate the product of links on the plaquette to find the flux
through the plaquette. 
  One thus dispenses with the simulated Higgs field.
Either the discrete-Higgs or the lattice-gauge-theory method results in the assignment of $S_3$ elements to each link of the lattice, and one must only
evaluate a plaquette to determine whether it is pierced by a string.
  These two methods give rise to string distributions which may
differ in detail but are qualitatively similar.  Experience with the $Z_3$ model (including my own simulations) shows that the subsequent evolution of the network is not sensitive to the details of the initial conditions:  $Z_3$ networks generated by the discrete-Higgs and lattice gauge methods begin to behave very similarly after just a few steps of
dynamical evolution.  We can test whether this is also the case in our $S_3$ model by comparing the evolution of networks generated by different methods.

\subsection{Initial location of nodes}

The lattice Monte Carlo algorithm, whether it is of the discrete-Higgs or lattice gauge type, generates a set of variables defined on a lattice.  From a lattice configuration, one can easily determine which plaquettes are pierced by strings, and it is also easy to determine the conjugacy class of the flux 
through a particular plaquette.  This, however, does not determine the precise location of each string inside the unit cube,  or to which other strings a given
string is connected.  Knowing lattice variables, we know through which cube faces strings emerge, but what happens to these strings inside the cube is 
completely unspecified.  Therefore, the next stage of our Monte Carlo algorithm must specify the actual locations of nodes within each lattice cube, and determine which nodes are connected to each other by straight string segments.  Having done this, we will finally be able to define all fluxes according to standard paths with respect to a single basepoint.

In generating a configuration of nodes and strings within a particular unit cube, we use the following criteria:
1)  The configuration within a cube should not be more complicated than necessary.
2)  Every string which pierces a cube face must end at a node inside that cube.
The reasons for this stipulation should become apparent later; one is that
it ensures that the flux of the string will be well-defined at some point inside the cube (since our conventions define string fluxes near their ends).
3)  Nodes inside a cube are placed approximately (not necessarily exactly) at the center of the cube.  This is purely for the sake of simplicity.
4)  Nodes inside the same cube should be separated from one another by some non-zero distance.  In order to define the conventional clockwise order of strings at each node, it is necessary that all strings have nonzero length.  It is for this reason that not all nodes are placed {\it exactly} at the center of
the cube.
 
Anywhere from two to six string ends, or none at all, may emerge from a given unit cube.  A cube with
three ends emerging is simple:  we simply place a single vertex at the center of the cube.  If exactly two plaquettes of a cube are nontrivial, we could interpret this as a single string entering the cube through one face and exiting through the other.
In compliance with criterion (2) above, our algorithm interrupts such a segment with a doubly linked pair of nodes.  This ensures, as previously mentioned, that the flux of the string will be well-defined at a point inside the cube (a necessity
when we perform the gauge fixing in order to define all fluxes with respect to the same basepoint.)  It also allows for the possibility that the string bends
inside the cube.  A cube with string ends emerging from two adjacent faces implies that the string bends as it passes through the cube.  Since our simulation only allows for straight string segments between nodes, a bent string
can only be implemented by introducing a doubly linked pair of nodes.  In the later dynamical evolution of the network, these nodes may annihilate, allowing the string to straighten.  In the process of straightening, the string may be obstructed by other non-commuting strings, and so it is important that the straightening not be allowed to happen until all string fluxes are suitably well-defined to allow the necessary comparison.  In order to separate the two
nodes by a finite distance (see point (4) above), each one is displaced slightly away from the center and toward the cube face through which its string emerges.  Examples are shown 
in figure \FIG\twoends ~\twoends.

\insertxfigpage{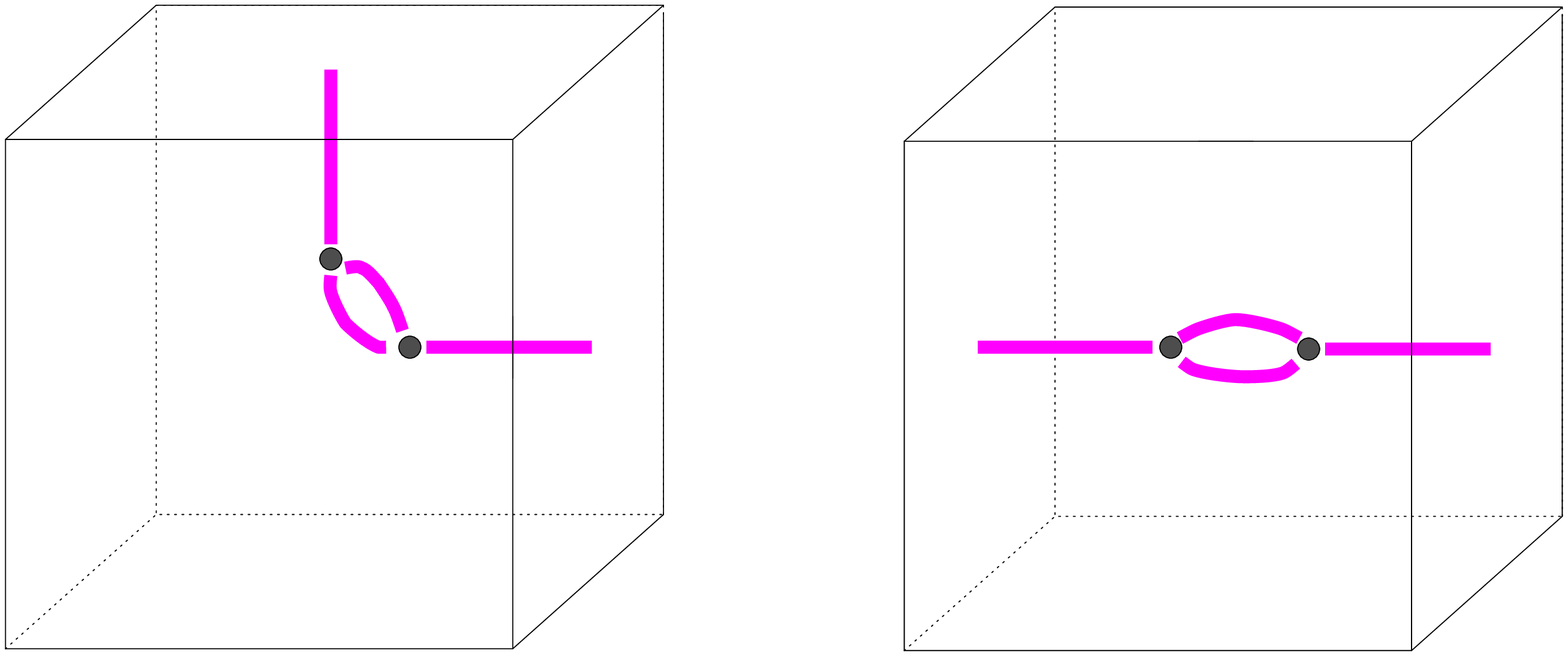}{\bcap\twoends \/ When strings pass through exactly two ends of a lattice cube, this represents a segment, either straight or bent, passing through the cube.  Our Monte Carlo algorithm represents this as a doubly linked pair of nodes.}

Cubes with four or more emerging strings are more complicated.  In general, there may be more than one consistent way of connecting the ends inside the
cube.  In our algorithm, we attempt to choose more or less at random from among 
the set of possibilities.  A helpful observation is the following:  If two plaquettes are nontrivial, but the product of links along a path which circumnavigates both is trivial, then it is possible to connect the two associated string ends through a
string segment which has no linking with any of the other strings in the cube.
(See figure \FIG\freeseg ~\freeseg.)  Let us call this a ``free segment.'' 
Our procedures for dealing with cubes of four, five ,or six emerging strings 
consist, roughly speaking, of searching in random order for a pair of faces allowing such a free segment.  If one is found, then it is formed in the same manner as for a cube with two ends, i.e., by inserting a doubly linked pair of nodes.
In the case of four or five ends, the connection of the remaining ends is determined
as soon as one free segment has been established.
In the 4-end case , the two remaining ends join in an another free segment; in case of five, the three remaining ends meet at a node.  In the case of a cube with six outgoing strings, if a free segment is produced from two of the ends, then a similar search for a consistent configuration can be performed on the remaining 
four ends.   For a cube with five outgoing ends,  another possibility, aside from one free segment and one node, is a 3-node configuration of the type shown in figure
\FIG\hungseg ~\hungseg,  in which one of the nodes is attached to just one outgoing string, and the others are each attached to two outgoing strings.

  The procedure of our Monte Carlo algorithm for establishing the nodes inside a cube with five ends is to search through a list of possible patterns involving either
one free segment or none, starting at a random place in the list and checking them for consistency by evaluating appropriate products of links around pairs of
cube faces.  When a pattern is found which is consistent, the nodes are established in that pattern.  

For a 6-string cube with some $t$-strings, the procedure as follows:  Pick, at random, two adjacent faces of the cube, and determine whether or not those two faces can be connected by a free segment.  
If so, connect them; otherwise,  they are assumed to join at a node which has
an additional connection to another node inside the same cube.  Then pick randomly a pair of adjacent faces from among the remaining four and apply a similar procedure.  The possible results of this procedure are that zero, one, or three pairs of adjacent faces are joined by free segments.  If only one pair
is connected by a free segment, then the other four strings are connected in pairs to nodes which are then connected to each other 
(figure \FIG\sixends ~\sixends B).  If no free segments are found, then there are three nodes, each attached to two of the outgoing segments, and we connect these by additional string segments to a fourth node located at the center of the cube (figure \sixends C). 

\insertxfigpage{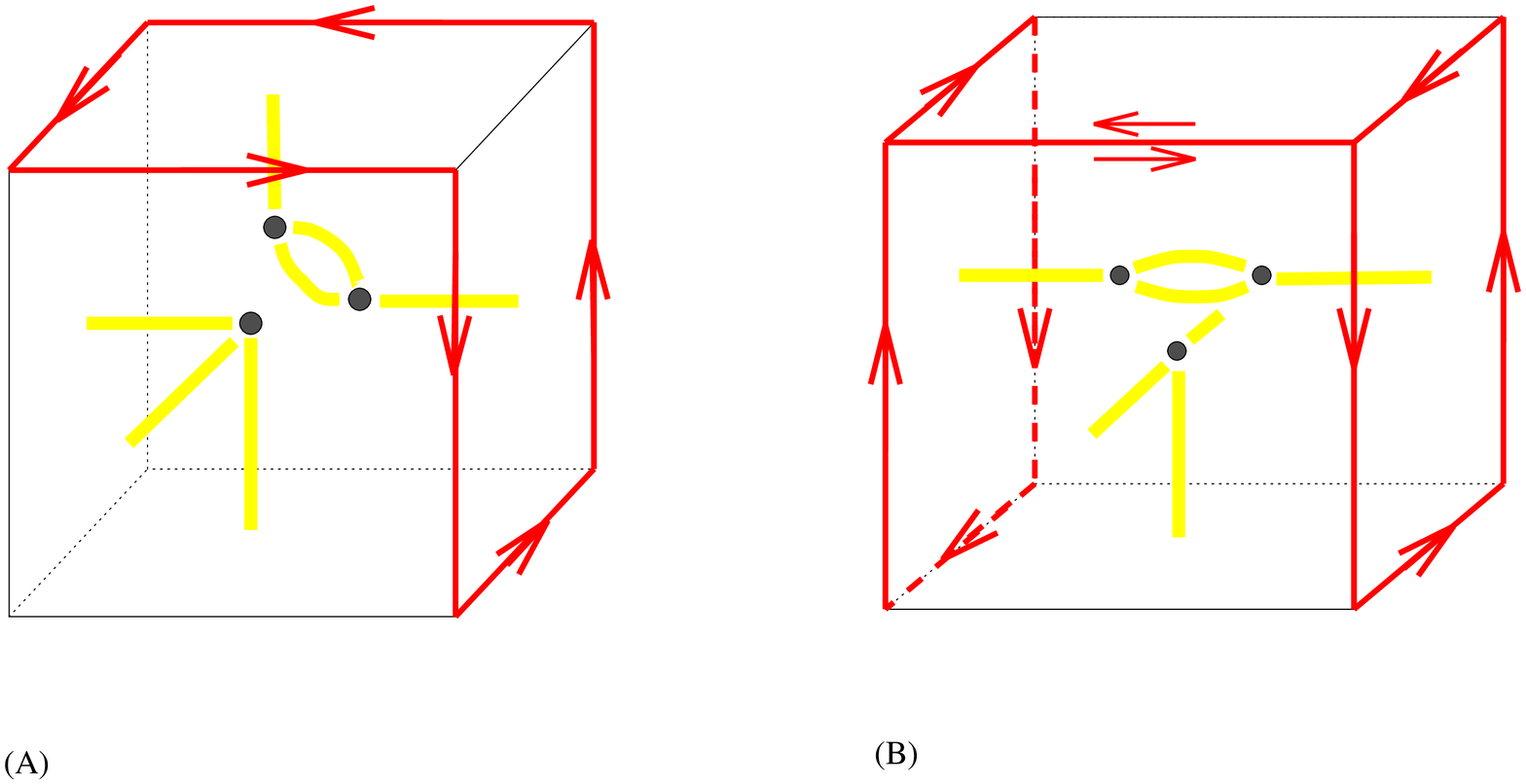}{\bcap\freeseg \/ (A) If a string segment (or a string
segment interrupted by a doubly linked pair of nodes)  passes into a cube through one face and out through an adjacent one, with no linkage to any other strings in the cube as shown here, then the highlighted path along the edges of both faces can be continuously shrunk to a point without crossing any strings; hence the corresponding product of links must be trivial.  Conversely, if the product is not trivial, then the connection of those two string ends by a free segment is not consistent.  An analogous criterion applies in (B). }

\insertfigpage{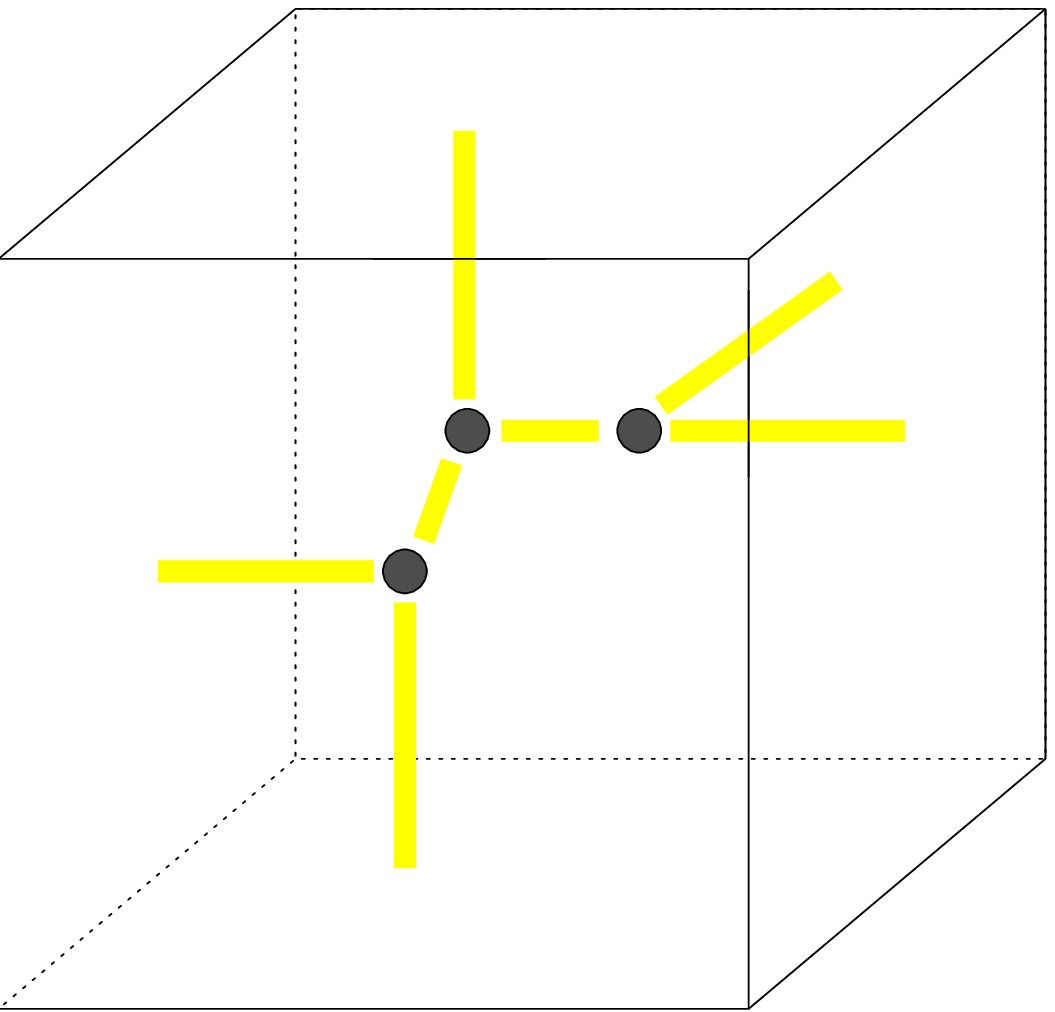}{\bcap\hungseg \/ Example of a cube with five
 outgoing strings but no free segments.}

\insertfigpage{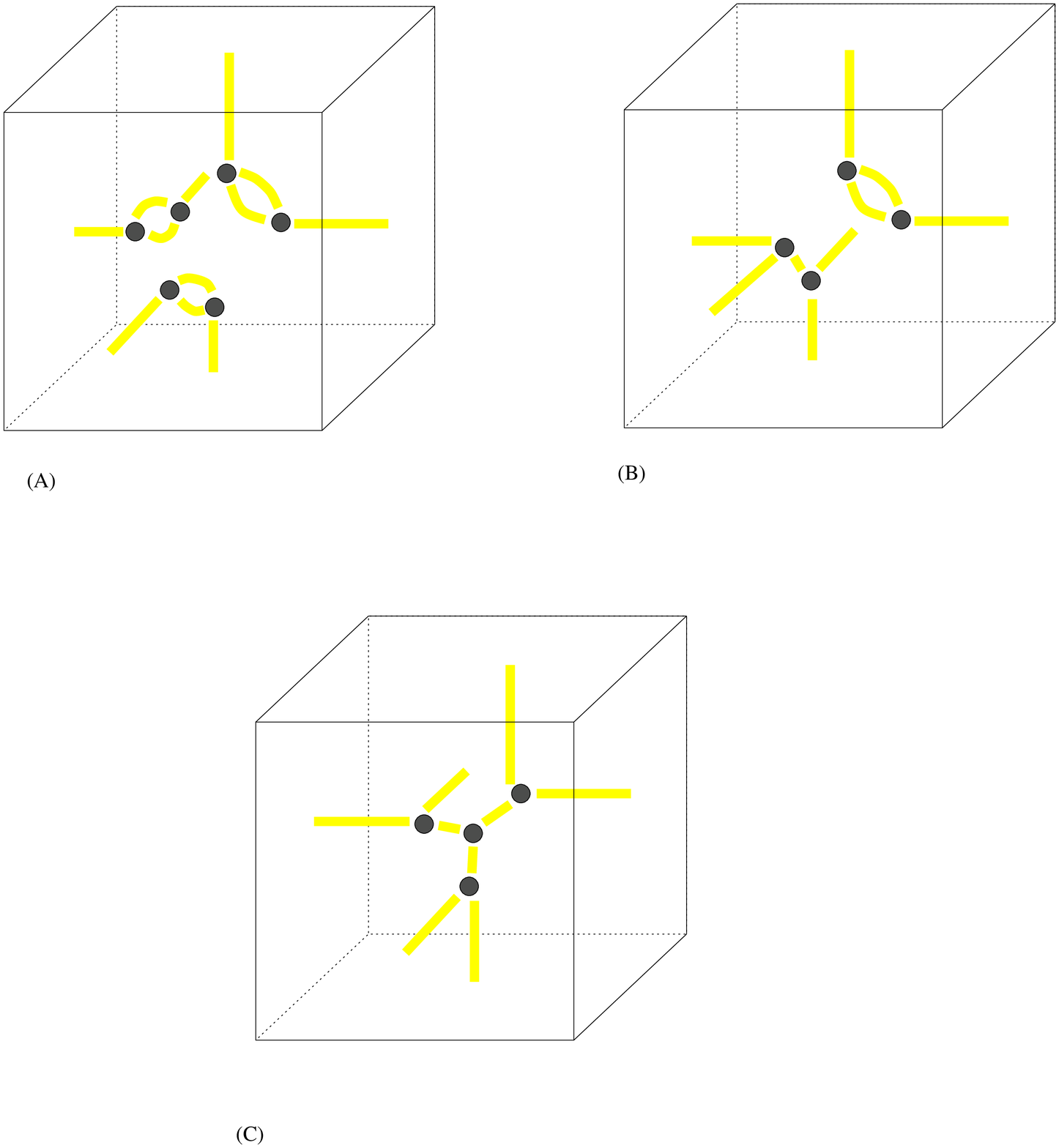}{\bcap\sixends \/  Configurations for a cube with all six faces pierced by strings:  (A) Three free segments.  (B) One free segment.
(C)  No free segments.  }

In order to ensure a nonvanishing distance between nodes, we displace each node 
by a small distance (such as .05 of the lattice spacing) from the cube center in the direction of each external string attached to it.  For example, if a node is connected to strings which exit the cube in the $\hat{x}$ and $\hat{z}$ directions, the node is displaced
in the $\hat{x} + \hat{y}$ direction.  Finally, a small random perturbation is
added to the position of the node.  This last step is purely a precaution against problems
which could arise in defining the fluxes if many nodes were precisely aligned along some ray from the basepoint.  

It is hoped that the above procedure generates an adequately representative initial configuration of strings and nodes.  Experience with the $Z_3$ model
indicates that the precise details of the initial conditions are not crucial.

\subsection{Gauge-fixing of the initial network}

When the location of all nodes has been established, along with their connections to each other,  the gauge fixing which establishes definitions of all of the fluxes must still be performed.  The problem is that initially, fluxes are defined only on paths consisting of lattice links.  The establishment of canonical flux definitions according to the prescription of the previous section requires that lattice paths be deformed into paths involving only straight-line tails from the basepoint and short loops encircling the strings near their ends.  We proceed in two steps.  First, we perform a gauge fixing within each individual unit cube, choosing one corner of the cube as a local basepoint and defining a tail from this basepoint to each of the nodes within that cube.  Then these standard paths are attached to an external tail to form a path beginning and ending at the global basepoint.  Initially, this external tail consists of lattice links, but it can in turn be deformed to a straight line path. 
 
The first step, local gauge-fixing on a lattice cube, is illustrated in figure \FIG\cubefix ~\cubefix.  Each string end emerging from a cube is treated individually.  We first compute, by multiplying link variables, the flux through a lattice path which encircles the string in question.  If the string passes through one of the three cube faces which include the corner we have chosen as local basepoint, then this path is simply a plaquette; otherwise the path runs along an edge to the face which is pierced by string $f$, then around that face and back along the same edge, as shown in figure \cubefix A.  By construction, only one string may emerge through any given face.  Therefore, the lattice path can be shrunk to the one shown in figure \cubefix B without crossing any other strings, and thus this path must have the same flux.  The flux is now defined along a path consisting of a tail and a small loop enclosing only the string, near the point where it pierces the cube face.  We then wish to straighten the tail as in figure \cubefix C, and finally slide the path so that  it encircles the string near its
end (figure \cubefix D) as do the standard paths of figure \tail .  In some cases these last two steps can occur without any other string segments being crossed,  and so the flux as defined in figure \cubefix D is the same as that of the original lattice path in \cubefix A.  In other cases, however, the tail may cross another
string as it is being deformed (figure \FIG\cubeobstruct ~\cubeobstruct).  Then the flux of the first string must be conjugated by the flux of the other.  Hence, the flux of the first string cannot be defined until that of the other string has already been defined.  When computing the fluxes of the strings emerging from a given cube, we first deal with those ends which do not depend on other strings, and then compute others.  As long as every string passing through
a cube ends at a node inside the cube, it will not be necessary at this stage
to refer to any nodes outside of the cube.

\insertfigpage{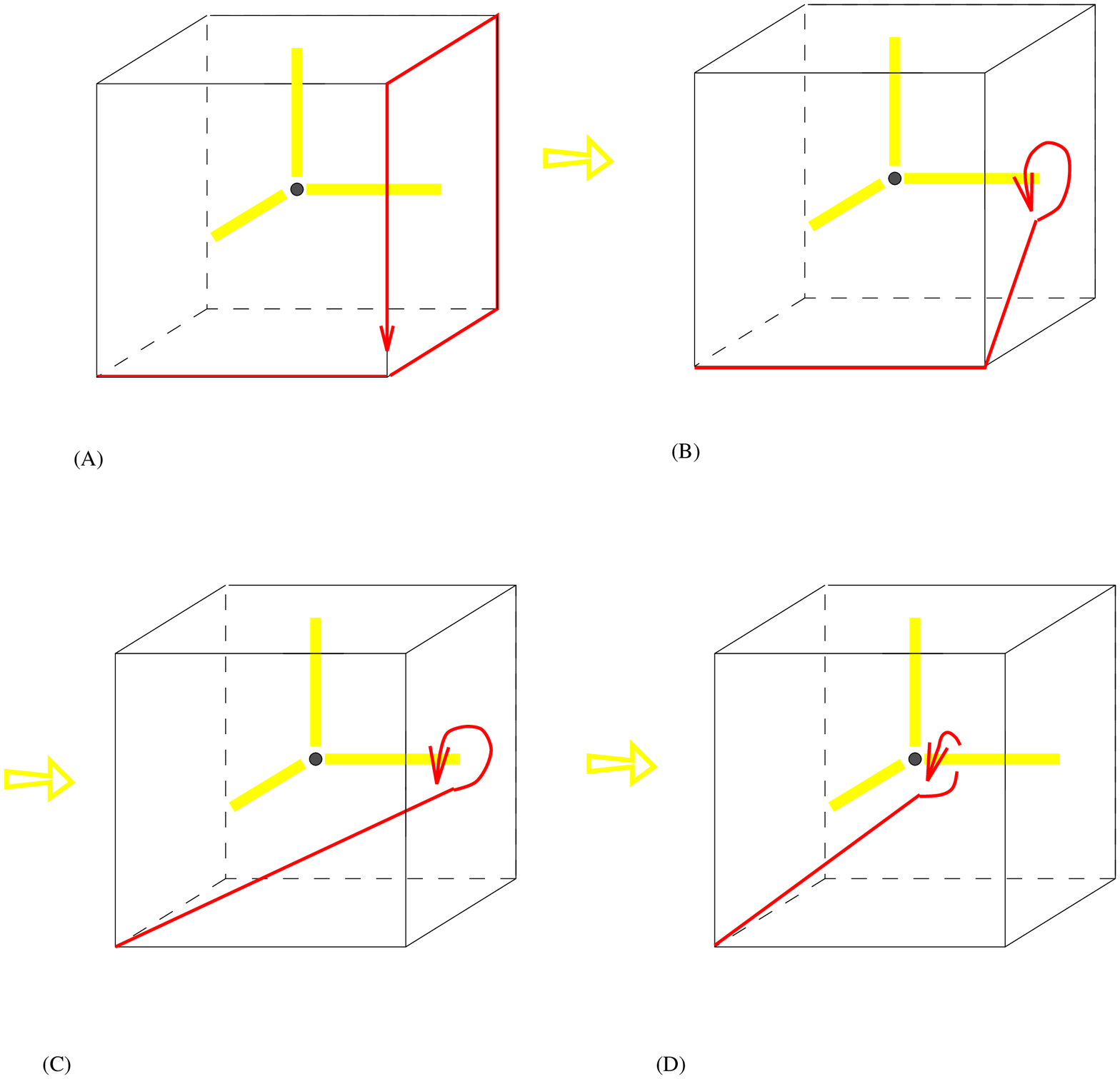}{\bcap\cubefix \/ ``Local'' gauge-fixing on a unit
cube of the lattice.  The flux of a string piercing the cube face is defined
first along a path consisting of lattice links.  The path is then deformed to
a path which runs along a straight tail to the node inside the cube, and encircles the string near its end.  If this deformation can be done without
crossing any other strings, then the flux defined by the final path shown in
\cubefix D  is the same as that defined by the lattice path in \cubefix A.}

\insertxfigpage{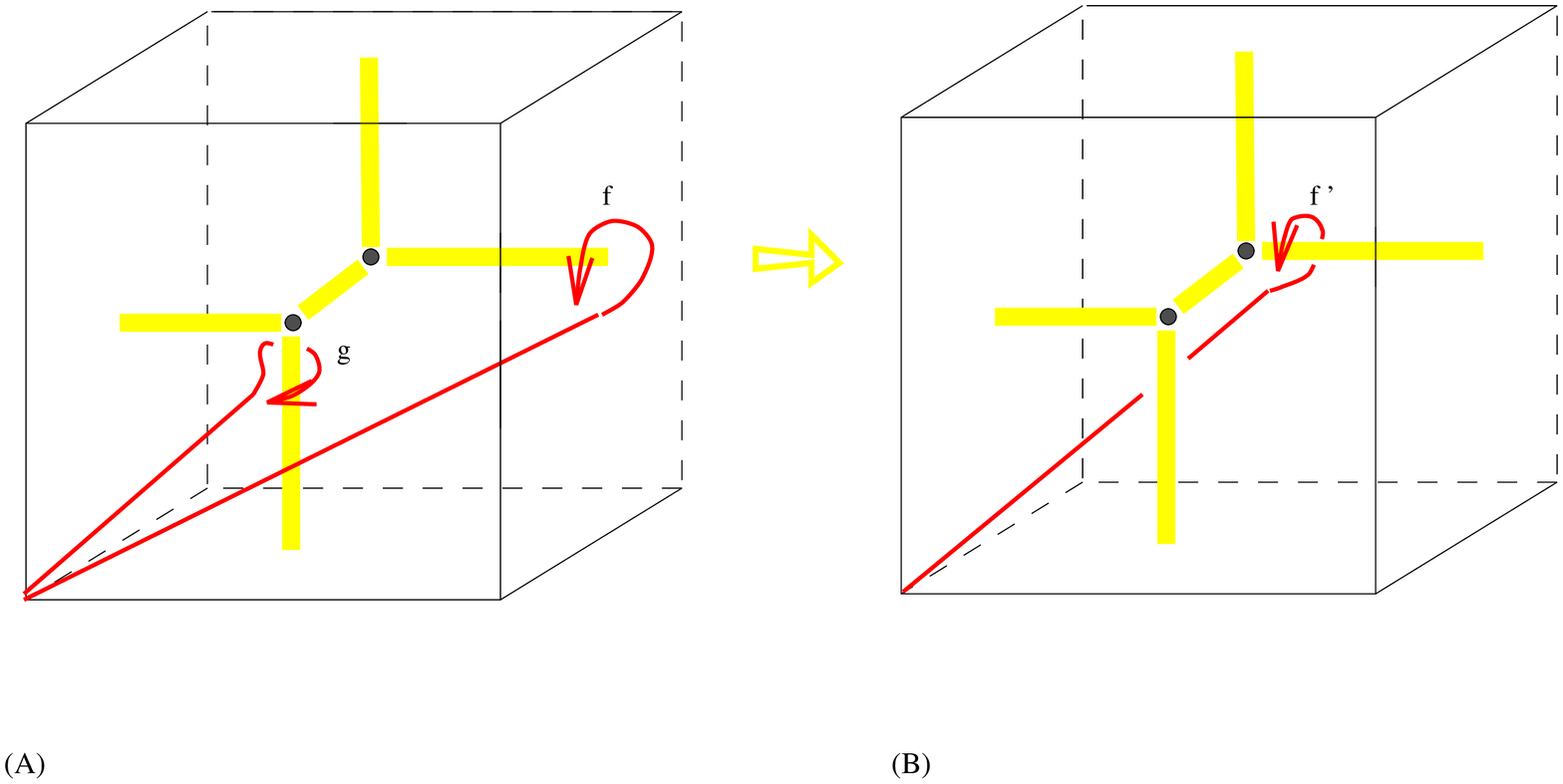}{\bcap\cubeobstruct\/  Example of an obstruction
to the last stage of the deformation shown in figure \cubefix:  A string has
flux $f$ as defined by the path in \cubeobstruct A (which corresponds to
\cubefix C),  but in the final stage of deformation, the tail crosses another string
whose flux is $g$.  The flux $f$ must be conjugated:  in this case 
$f'=g^{-1} f g$.}

In order to complete the definition of flux, we attach the path from the cube corner to the node to a tail which runs from the global base point to the cube corner.   This tail is initially defined as lattice path of the type shown in figure \FIG\latticeTail ~\latticeTail, consisting of three straight segments along each of which only one coordinate changes.  If the product of links along this tail is $t$, and
the ``local'' flux of a path within the cube starting and ending at the corner
is $a$,
then the new flux of the string is $t^{-1}at$.  The tail is then deformed to a single straight line by several steps, as shown.  At each stage of the straightening, other strings may be crossed within other cubes.  The flux measurement is conjugated by the appropriate fluxes.  Finally, we arrive at a flux defined along a path which proceeds along a straight line to the corner of
the unit cube in question and then from there to a node, where it encircles string $a$.  As a final step, we straighten this path to a single line segment
(figure \FIG\gridtaillast ~\gridtaillast), obtaining a definition of the string's flux according to the conventions of 
section 2.  Since this definition depends on the values of string fluxes in other cubes,  the unit cubes of our lattice must be handled carefully in order.  The flux definition procedure must be applied first to those cubes closest to the basepoint, and then, layer by layer, to the cubes farther away.

\insertxfigpage{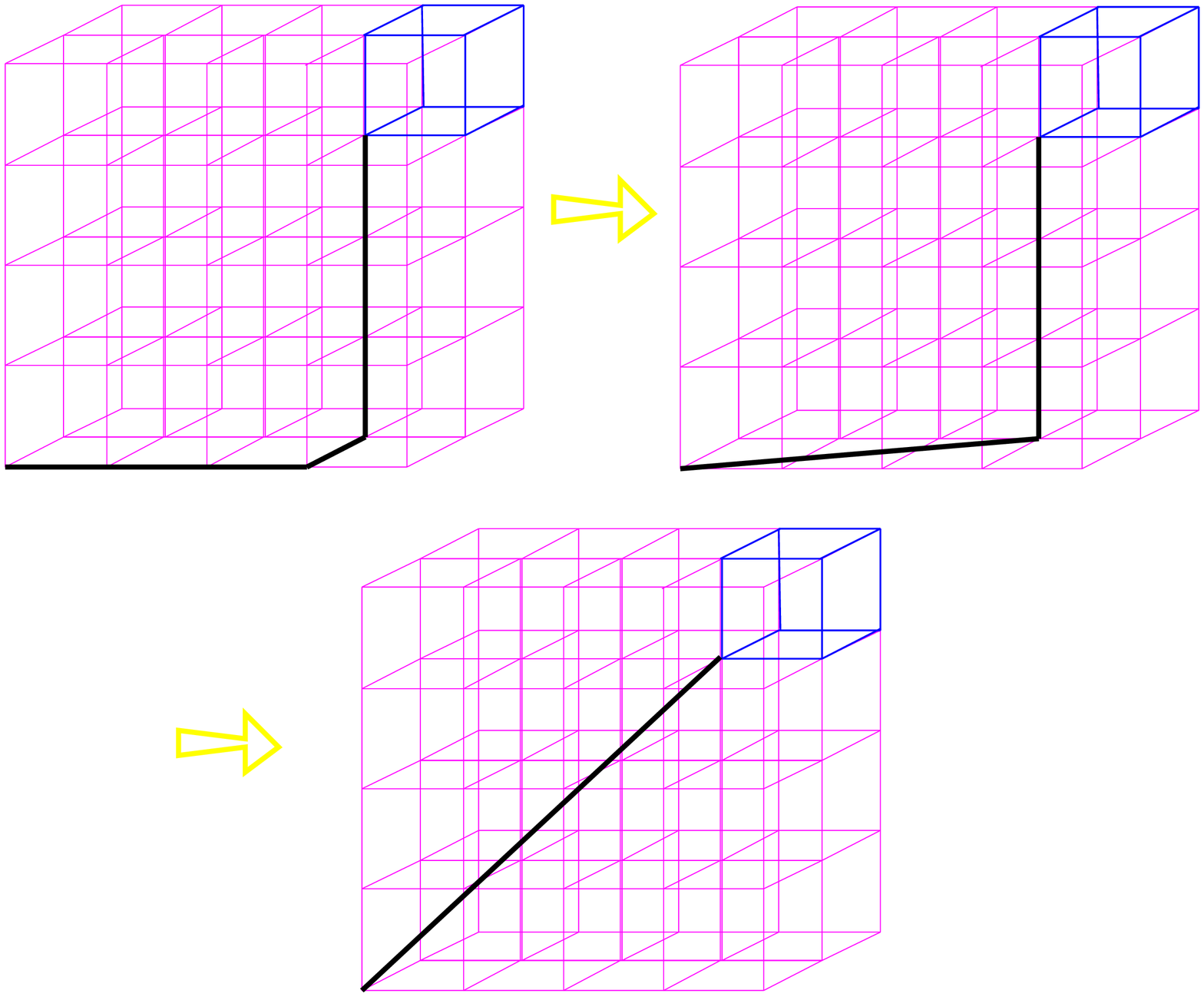}{\bcap\latticeTail\/  Defining a tail from the 
basepoint to the corner of a lattice cube.  Initially, the tail is defined as
a product of lattice links.  This tail is then deformed to a single straight
segment.  If other strings are crossed during this deformation, then the flux
is adjusted appropriately.  This procedure requires that the fluxes of strings
in intervening cubes have been defined already.}

\insertxfigpage{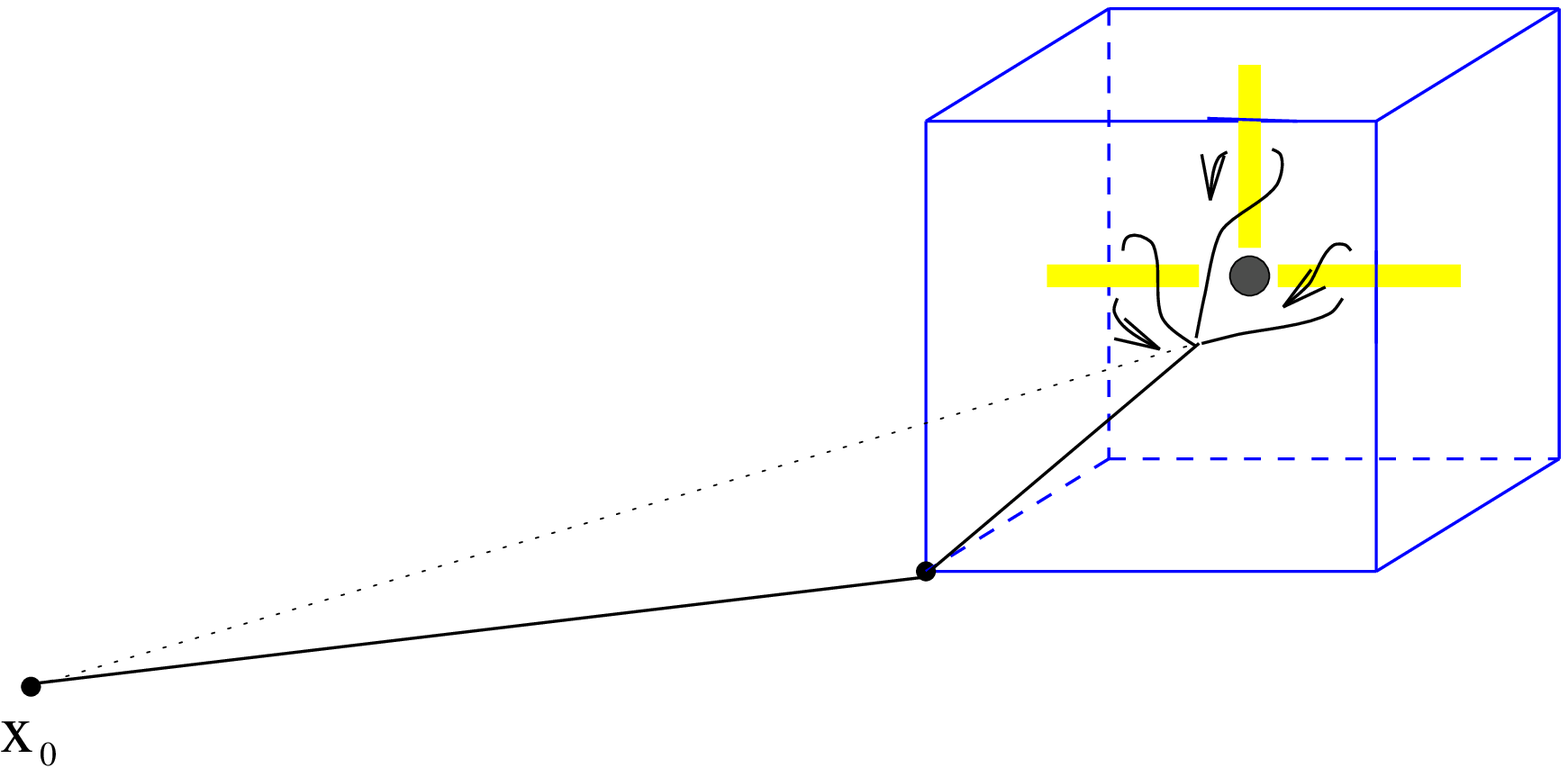}{\bcap\gridtaillast \/ When the procedure of
figure \latticeTail\/ has been applied, the tail from $x_0$  to a given node
consists of one straight segment to the corner of a unit cube, and another to
the node itself.  The final stage is to remove the last kink, straightening this
tail to the one shown here as a dotted line.  Then all fluxes of strings joining
at the node are well-defined according to the conventions used in the dynamical
simulation.}

 After the fluxes of all strings emerging from the cube have been defined,  those which link one node to another {\it inside} the same cube may be fixed by means of the flux conservation condition:  the product of all strings
joining at a node (taken in clockwise order as defined in the previous section) must be the identity.  With all string fluxes thus defined, and the fluxes of the wrap-around paths $\Gamma_i$ computed by multiplying lattice links, we then have 
a fully specified initial condition for the evolution of the network.

\section{Dynamical Evolution}
After the establishment of initial conditions,  We may proceed with the dynamical
evolution of the network.  The system we are modelling, 
as in ref. \VV, is one of monopoles (vertices)  connected by cosmic strings, which we are modelling as straight line segments.  The monopoles,
as stated earlier, are assumed to undergo damped motion under the influence of string tensions.    Our simulation proceeds as follows:
During each time step, each node is moved by a displacement proportional to the vector sum of all tensions acting on it.  This type of evolution corresponds to damped motion: ${\rm Force} \prop {\rm Velocity}$. The constant of proportionality is a parameter which may be absorbed into the size of the time step.  However, the ratio of damping constants for the two types of nodes may
 be a separately adjustable parameter.
The ratio of tensions of the two different types of strings ($s$-\/ and $t$-type)
is also separately adjustable.    The nodes are moved one at a time during each time step.  During the motion of a given node,  all effects of this motion on the flux definitions throughout the space are monitored.  (For example, the necessary adjustments are made if one of the moving node's strings crosses the tail of another.)  

In addition to this simple motion of the node,  the following other types of events may occur.

i)  Intercommutation.  If in the process of moving a node from its initial to
final position, one of its string segments intersects some other segment, then the fluxes of those two segments are compared at the point where the crossing occurs.  If these two fluxes commute, then the the two segments may either pass through each other unaltered, or intercommute.  The probabilities of these two outcomes may be taken as an adjustable parameter of the simulation.  It is widely believed that intercommutation is generically the more common outcome whenever two cosmic strings cross.  Thus it seems most natural to let the probability of intercommutation be 1, or close to 1.  Intercommutation may occur in two possible situations:  either both strings are 3-cycle strings, or both are 2-cycles.  In the latter case, the fluxes of the two strings must in fact be equal.  In an intercommutation, the string ends are rearranged in such a way as to conserve flux.  In the case of two $s$-strings, there is always only one way to rearrange the ends, as shown in figure \FIG\intercom ~\intercom a.  A string end carrying flux $s_+$ to the point of intersection may not be joined to one carrying the inverse flux $s_-$.    When two $t$-strings intercommute, however, there are two possible rearrangements of the ends, owing to the fact that a 2-cycle is equal to its own inverse and 2-cycle or t-strings consequently have no preferred orientation{\footnote{*}{Strictly speaking, we can only say that there is no 
{\it topological} reason for a t-string to have a preferred orientation.  It is possible that the field equations could have two distinct solutions, corresponding to differently oriented strings, which are topologically equivalent but can be deformed into one another only by surmounting a finite energy barrier.  A situation of this sort occurs in the global vortices of nematic liquid crystals.  This was pointed out to me by J. Preskill.}} (fig. \intercom b.)  In the absence of a reason to prefer one of these rearrangements over the other, the choice must be made randomly.  After intercommutation, the newly joined segments straighten immediately.  Of course, all necessary adjustments are made if they should cross any tails during the process of straightening.  It may also happen that as the rejoined segments straighten, they intersect other segments, which may lead to further intercommutations.  Segments straighten as much as they are able to before encountering obstructions.

\insertfigpage{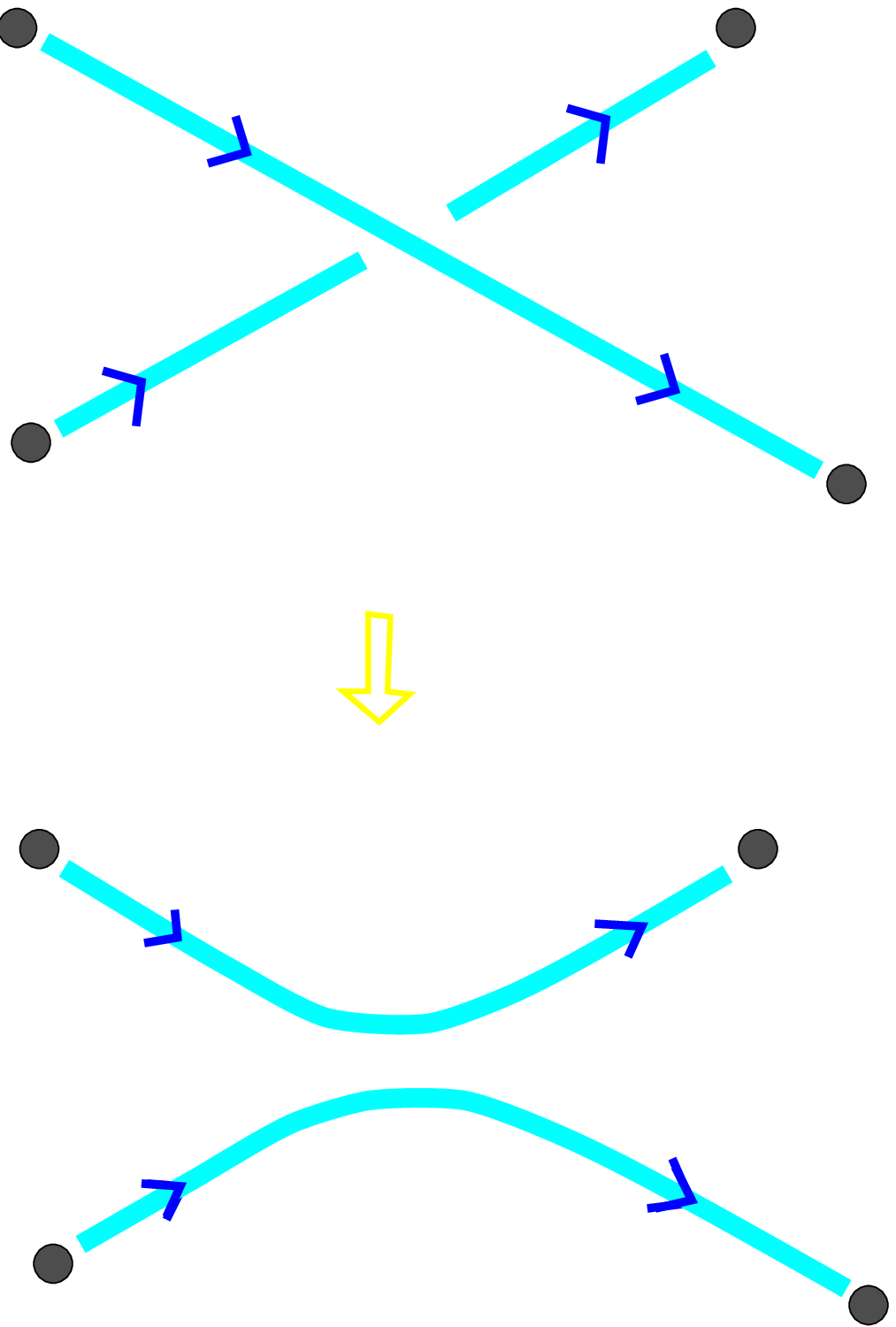}{\bcap{\intercom A}\/  When two three-cycle or $s$-strings intercommute, there is a unique rearrangement which is compatible
with the orientations of the strings.}

\insertxfigpage{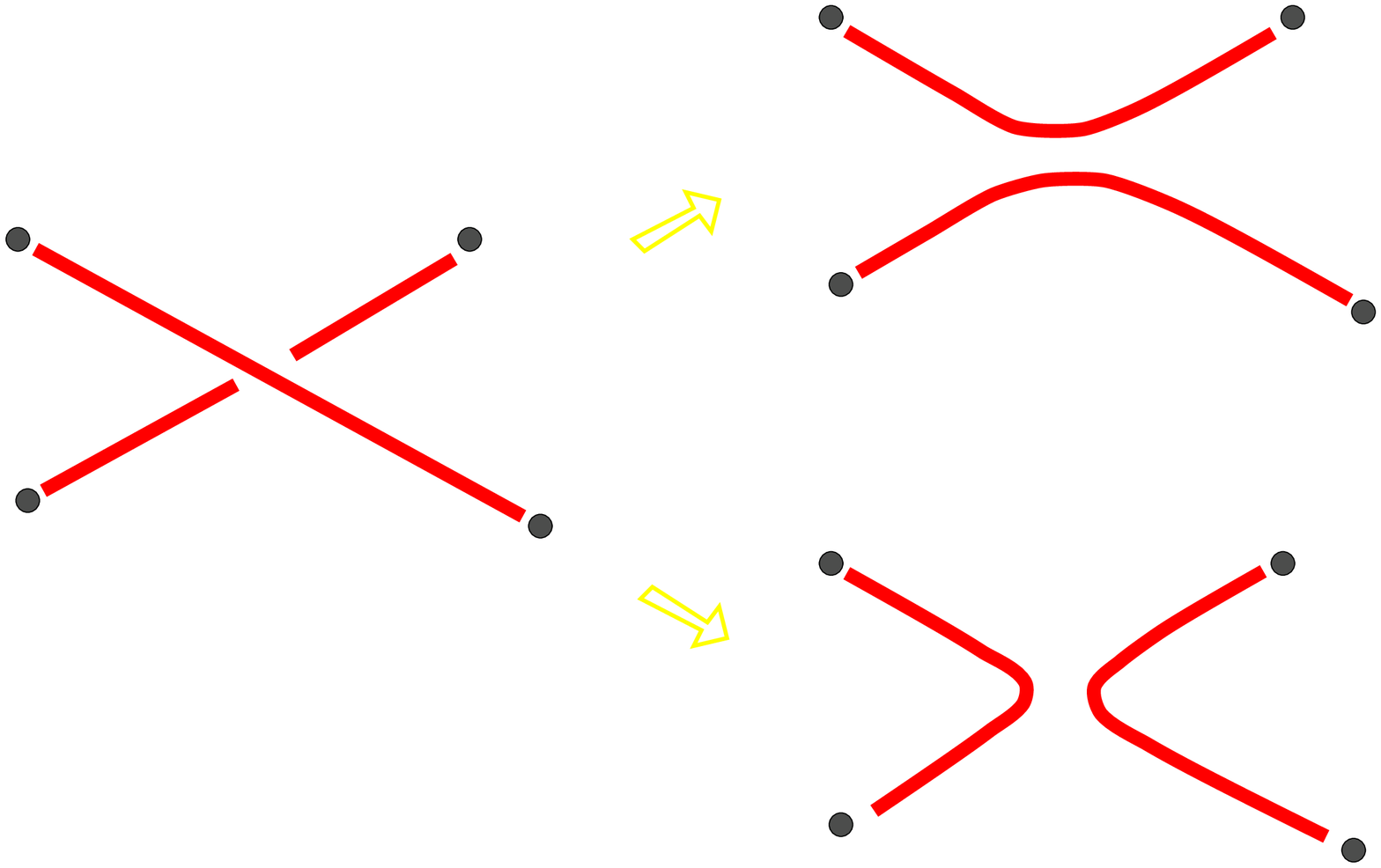}{\bcap{\intercom B}\/  Since $t$-strings (2-cycle
strings) have no orientation, an intercommutation can result in either of two possible
rejoinings of the ends.}

ii)  Non-commutative intersection (or linking).  If two non-commuting strings intersect, then it is assumed that they form a new pair of nodes and thus become linked by a new segment, as shown in figure \FIG\NCI ~\NCI.      The flux of the intervening string segment is uniquely determined by the requirement of flux conservation.
(The intervening flux must always be a 3-cycle, as the commutator subgroup of $S_3$ is $Z_3$.) 

The actual implementation of this linking process in the simulation is slightly tricky.   Defining the flux measurements of the two new nodes requires a
reference to previously existing nodes.  But linking events occur when one node is moving, dragging an attached string segment to cross another.  One must be careful that this motion not cause holonomy-type interactions with the newly created nodes before the new nodes are properly specified in terms of their associated fluxes.  Our strategy is to create and fully specify the new nodes 
first, before  the previously existing node is moved at all.  This requires that the
linking of strings actually be accomplished before the existing node has moved.
A prescription for doing this is shown in figure \FIG\NCIDetail ~\NCIDetail.
Related issues arise to some extent in implementing an intercommutation event: intercommutations occur when one node is moved, dragging along a string segment
which intersects another as is it is being dragged.  Tail crossings and other
holonomy interactions occur both as a result of the node's motion and as a result of the rearrangement of string ends during the intercommutaion,  the non-commutativity of fluxes requires careful attention to the order in which these interactions are handled.  As in linking events, the present simulation follows a strategy of making all rearrangemements of string ends before any existing nodes are actually moved. 

The linking process creates new nodes and new strings, and therefore it might be expected to impede the collapse of the network.

\insertfigpage{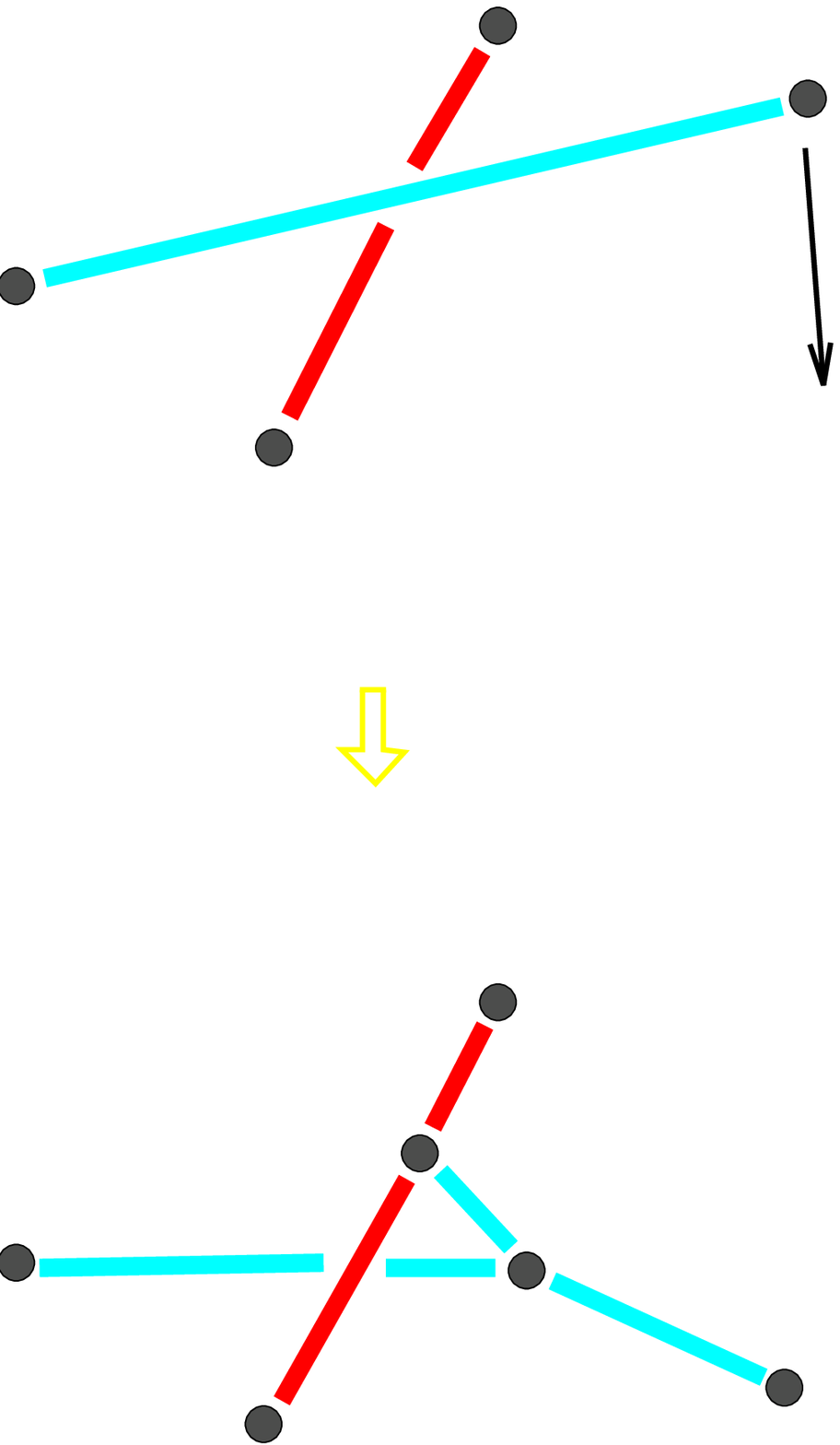}{\bcap\NCI \/ The intersection of two strings whose fluxes
do not commute causes them to become linked by a new segment.}

\insertsmallxfigpage{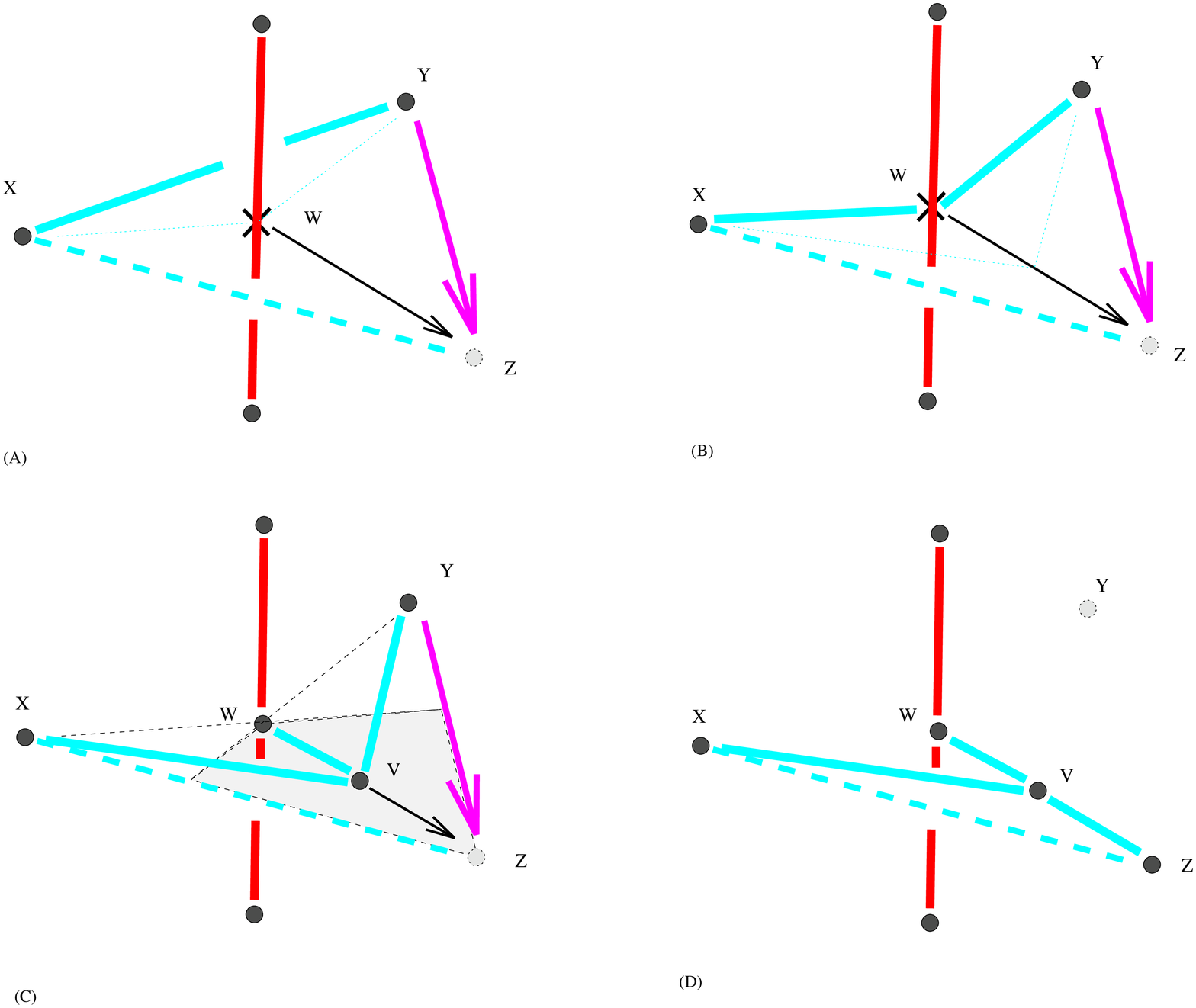}{\bcap\NCIDetail \/ Actual implementation of the ``linking'' or ``non-commutative intersection'' process shown in Fig. \NCI \/.
\/(A)  If the node at point $Y$ were to move to $Z$ while dragging with it the
string segment $XY$,  the string would intersect another one (with non-commuting flux) at the point labelled $W$ and marked with a cross.  We create the new nodes first before moving the existing node. \/ (B) First, the two nodes are both created at the intersection point $W$, and the moving segment is deformed
to two segments joined at $W$, making all necessary adjustments for the crossing of other nodes' tails, etc.\/  (C) In order to define the conventional
clockwise order of the two new nodes, they must be separated.  We do this by moving one of them to a new point $V$.   
$V$ must be chosen so that the path consisting of $XV$ and $VY$ actually does lie on the opposite side of the other string as compared with
$XY$;  i.e.,  so that the linking of strings has occurred even though the node at $Y$ has not moved yet.  Assuming $V$ is coplanar with $X$,$Y$ and $Z$, this means it must lie in the shaded region.  This can be ensured by placing $V$ somewhere along the segment $WZ$, and  we may arbitrarily choose to put it at the midpoint of that segment. The fluxes of the new nodes at $W$ and $V$ are then 
fixed by requiring consistency with the string segments to which they are attached and with the single-node consistency relation \fluxcons .\/ (D) Finally, the node may move from $Y$ to its destination $Z$.}

\FIG\doublelink

\insertfigpage{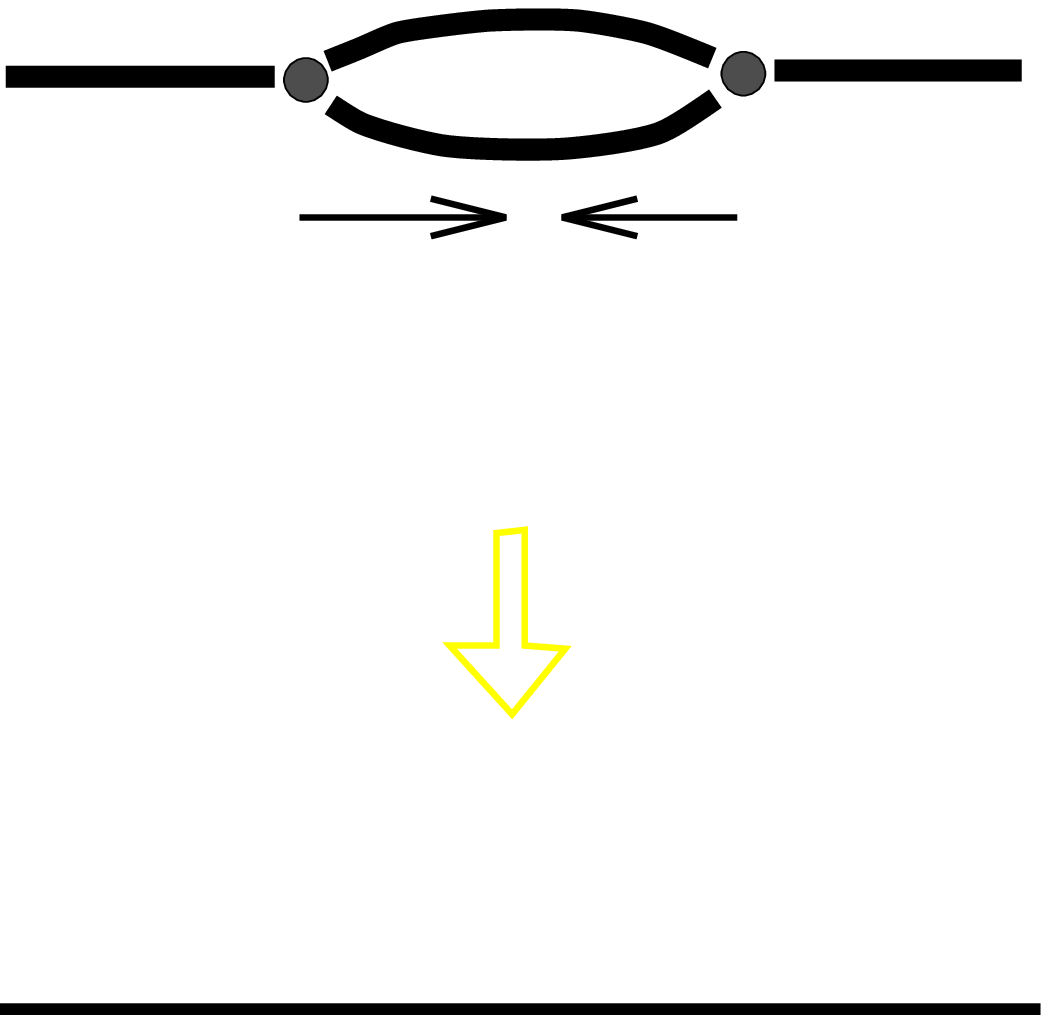}{\bcap\doublelink \/  A pair of nodes linked by
two strings may annihilate, leaving a single string.}

\FIG\sssannih

\insertxfigpage{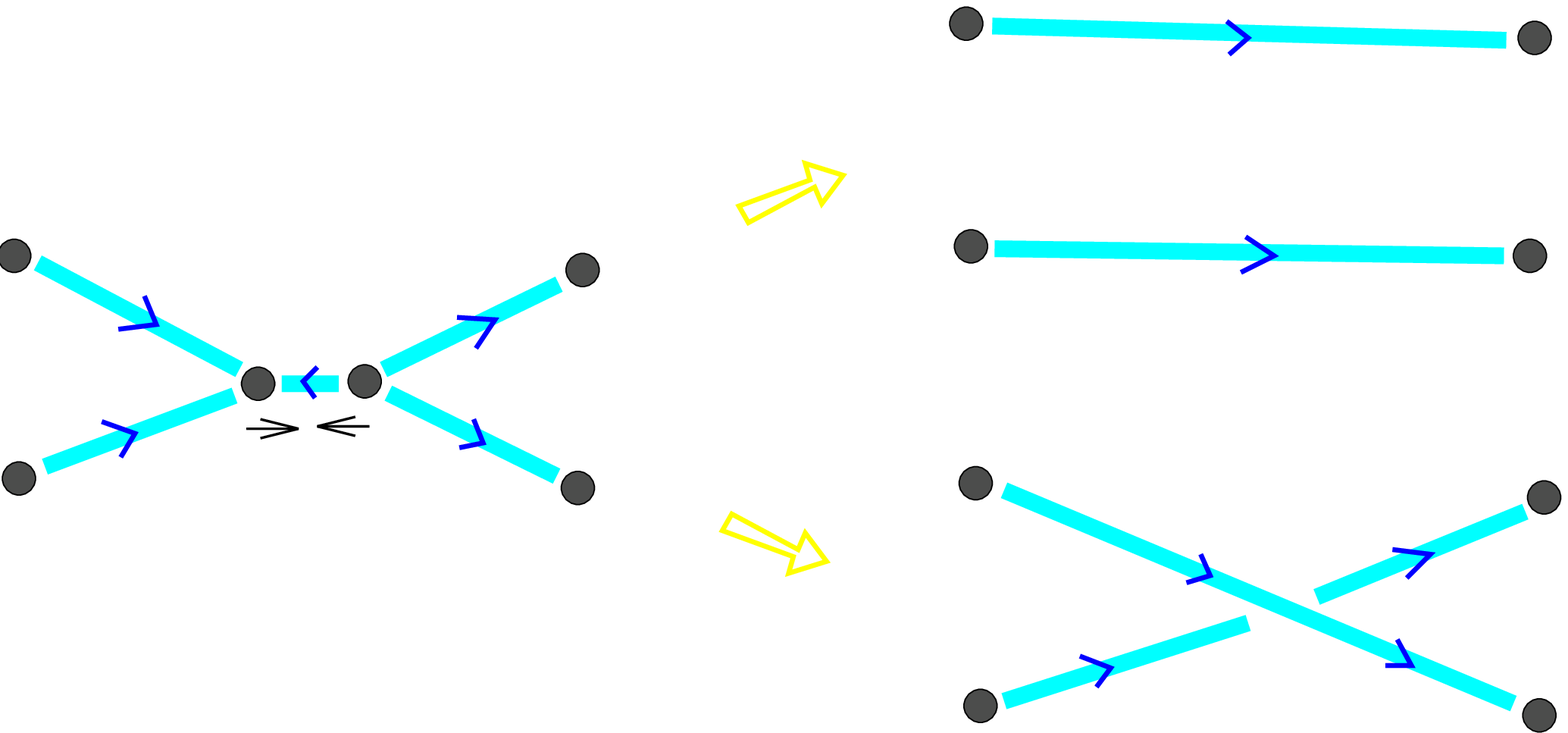}{\bcap\sssannih \/ Annihilation of two $sss$ nodes
joined by a single string.  There are two possible ways to reconnect the strings
consistent with their orientation.}

\FIG\cantannihilate

\insertxfigpage{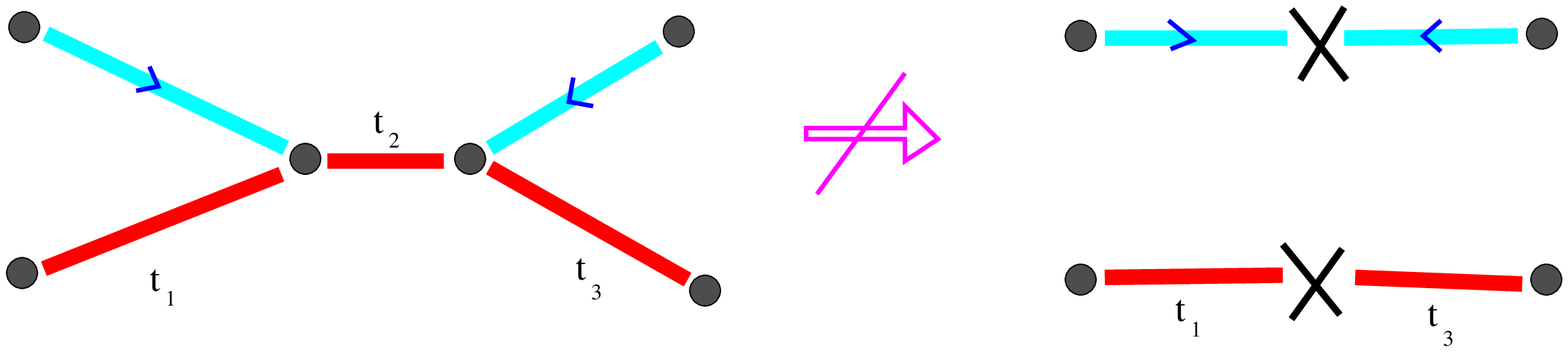}{\bcap\cantannihilate \/ The two nodes shown 
here cannot annihilate, because there is no consistent way to reconnect the 
string ends.}

 \FIG\indannih

\insertlongfigpage{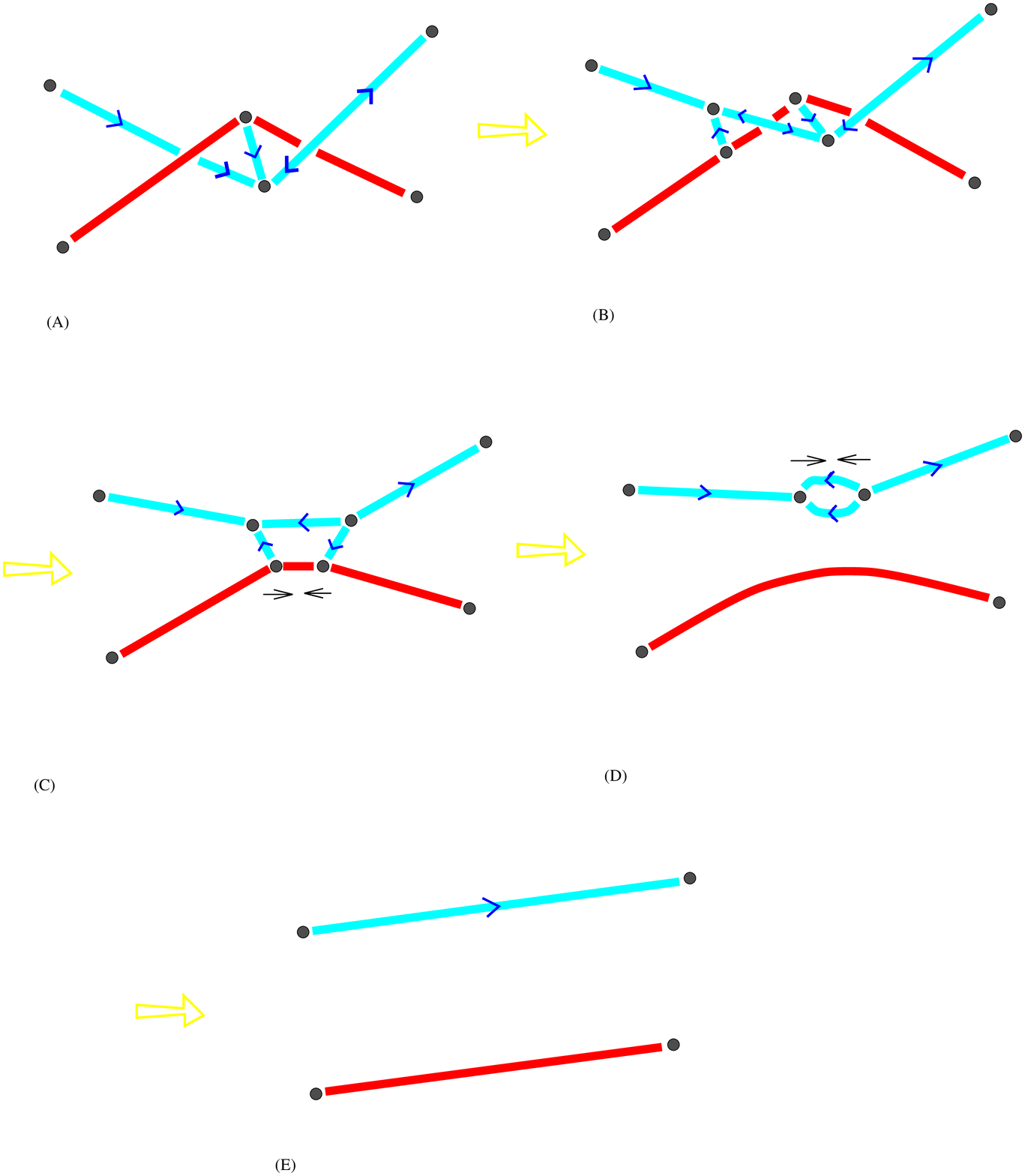}{\bcap\indannih\/ Unlinking of two strings-- the
inverse of the process shown in figure \NCI-- can occur in several steps if
the string tensions pull in the right direction to unlink the strings.  A
linking followed by two annihilations has the net result of removing the 
short intervening segment and unlinking the two longer strings.  In this 
figure, the basepoint is assumed to lie behind the page, so that the definition
of an $s$-string's flux changes when it passes in front of any $t$-string.}

iii)  Annihilation.  When two nodes approach each other closer than a distance 
$r_{min}$ which is a parameter of the simulation,  they are allowed to annihilate.  The segment(s) which join the two nodes is eliminated, and the other segments emanating from the two annihilating nodes are joined to each other and straigtened (or straightened until an obstruction is encountered). 
Two nodes are able to annihilate only if there is a consistent way to rearrange the free string ends (i.e., each string is able to find a partner with the same flux).  Annihilation is always possible if the two nodes are doubly 
linked as shown in figure   ~\doublelink.  It is also always possible if both junctions are of the sss type, even if they are only singly linked.  In this case, there are two possible rearrangements of the free string ends (figure   ~\sssannih).  One of these two must be chosen at random.  When two stt-type junctions approach each other, there may  exist
at most one consistent rearrangement of the free ends allowing the two nodes to annihilate.  Annihilation requires that {\it each} of the two segments on one side be matched with one on the other side carrying the {\it same flux}.  Figure   ~\cantannihilate\/ shows an example of a pair of nodes which cannot annihilate
because there is no consistent rejoining of the string ends.
If two junctions approach each other but are unable to annihilate, then they continue to move normally, going wherever are pulled by string tensions.  
They might either remain nearby or be pulled apart once again. If two stt-type 
nodes do annihilate, it is easily seen that there can never be more than one consistent rearrangement of string ends.  If two of the outgoing ends are $s$ strings and two are $t$ strings, then there cannot be more than one rearrangement because each string can only be joined with one in the same conjugacy class.  If all outgoing strings are of $t$ type, then all four cannot have the same flux-- if they did, then the total flux of any pair would be trivial and they would not be connected by a segment.  Nor may any three have the same flux.  It follows that, at best, each string end may reconnect with a unique partner.
  Annihilation of nodes is, of course, the principal mechanism by which the network dissipates its energy in our model.  

We might imagine another type of annihilation process which is the inverse process to the non-commuting intersection.  If string tensions were pulling in appropriate directions, then  two nodes might annihilate,  leaving two strings free which were previously connected by another string.   Because of, among 
other things, the difficulty of constructing an algorithm to determine when this may occur, we do not include such events explicitly in the simulation.   The  unlinking of two strings can occur, however, through a multi-step process involving several string intersection and annihilation events
(figure ~\indannih).
We expect that such a process probably {\it will} occur whenever the geometry is appropriate for the unlinking of two strings, so that it is not necessary to perform the unlinking ``by hand'' in a single step within the simulation.

\vskip 0.4in
 
\ack

I thank John Preskill for helpful discussions. This work was supported in 
part by U.S. Department of Energy Grant no.  DE-FG03-92-ER40701.

\par\penalty-400\vskip\chapterskip\spacecheck\referenceminspace
   \ifreferenceopen \Closeout\referencewrite \referenceopenfalse \fi
   \line{\fourteenrm\hfil REFERENCES\hfil}\vskip\headskip
   \input referenc.txa

\end